\RequirePackage{amsmath}
\PassOptionsToPackage{prologue,dvipsnames}{xcolor}
\documentclass[runningheads]{llncs}
\usepackage[T1]{fontenc}
\usepackage{graphicx}
\usepackage{booktabs}
\usepackage[misc]{ifsym}
\newcommand{\corr}{(\Letter)}
\usepackage{amsmath}

\usepackage{algorithm}
\usepackage{algpseudocode}
\usepackage{hyperref}

\usepackage{comment} 
\usepackage{makecell}
\usepackage{tikz}
\usepackage{amssymb}
\usepackage{subcaption} 
\usepackage{multirow}
\usepackage{mathtools}
\usepackage{diagbox} 

\newcommand{\crule}[3][black]{\textcolor{#1}{\rule{#2}{#3}}} 

\makeatletter
\newcommand*{\rom}[1]{\expandafter\@slowromancap\romannumeral #1@}
\makeatother

\newcommand\ib{\stackrel{\mathclap{\textbf{\_}}}{i}}

\usepackage{colortbl}
\usepackage{xcolor}
 \usepackage[autostyle=true]{csquotes}

\begin{document}

\title{Fairness is in the details : Face Dataset Auditing}

\titlerunning{Fairness is in the details}

\author{Valentin Lafargue\inst{1,2,3} \corr \and Emmanuelle Claeys \inst{4} \and Jean-Michel Loubes \inst{2,3} }

\authorrunning{V. Lafargue et al.}

\institute{IMT, Toulouse, France 
\email{valentin.lafargue@math.univ-toulouse.fr}
\and
INRIA, Toulouse, France
\and
ANITI 2, Toulouse, France
\and
IRIT, Toulouse, France
}

\toctitle{Fairness is in the details : Face Dataset Auditing}
\tocauthor{Valentin~Lafargue, Emmanuelle~Claeys, Jean-Michel~Loubes}

\maketitle              

\begin{abstract}
Auditing involves verifying the proper implementation of a given policy. As such, auditing is essential for ensuring compliance with the principles of fairness, equity, and transparency mandated by the European Union's AI Act. Moreover, biases present during the training phase of a learning system can persist in the modeling process and result in discrimination against certain subgroups of individuals when the model is deployed in production.
Assessing bias in image datasets is a particularly complex task, as it first requires a feature extraction step, then to consider the extraction's quality in the statistical tests.  This paper proposes a robust methodology for auditing image datasets based on so-called "sensitive" features, such as gender, age, and ethnicity. The proposed methodology consists of both a feature extraction phase and a statistical analysis phase. The first phase introduces a novel convolutional neural network (CNN) architecture specifically designed for extracting sensitive features with a limited number of manual annotations. The second phase compares the distributions of sensitive features across subgroups using a novel statistical test that accounts for the imprecision of the feature extraction model. Our pipeline constitutes a comprehensive and fully automated methodology for dataset auditing. We illustrate our approach using two manually annotated datasets. The code and datasets are available at 
\href{https://github.com/ValentinLafargue/FairnessDetails}{github.com/ValentinLafargue/FairnessDetails}.

\end{abstract}

\section{Introduction}


The widespread adoption of machine learning (ML) systems in industrial applications has heightened concerns about fairness, transparency, and accountability. 
The issue of bias in algorithmic decision-making has emerged as a critical concern within the machine learning community. A substantial body of research has examined how such biases can adversely impact algorithmic outcomes, potentially leading to violations of fundamental rights, as highlighted in the European AI Act. This legislation highlights the need to prevent AI systems from perpetuating or exacerbating existing societal inequities through systematic bias analysis. These biases not only compromise fairness but also raise ethical and legal challenges, underscoring the need for rigorous detection through systematic audit processes to ensure accountability and mitigate unintended harms. We refer for instance to \cite{fairbook}, \cite{barocas-hardt-narayanan}, \cite{10.1145/2783258.2783311}, \cite{besse2022survey}, \cite{risser2022tackling}, \cite{chouldechova2017fair}, \cite{gordaliza2019obtaining} or \cite{NIPS2016_9d268236}.  
Beyond decision-making contexts, we know that algorithmic biases often stem from biases present in the training datasets themselves. Auditing an image dataset  is a challenge in itself. Firstly, it is necessary to determine which variables to consider and how to extract them from an image. 
The importance of auditing image datasets is amplified by the fact that every image inherently encodes explicit features. Unlike text or numerical datasets, which can omit or abstract sensitive details, images visually represent specific characteristics, often revealing cues about sensitive features such as ethnicity, age, and gender.
Manual labeling of such features is prohibitively expensive when dealing with large-scale datasets or when conducting extensive audits across multiple variables. To address this issue, convolutional neural networks (CNNs) can be employed to predict sensitive features, although they require annotated data for their training (lesser amount). Once trained, the network can predict the sensitive feature of the remaining data in the dataset (with a certain error relative to it).
In our context, we define bias in an image dataset as statistically significative difference of distributions (e.g., an ethnicity or an age group under-represented). 
Statistical tests usually do not take into account the uncertainty of the labels (false predictions). 
We propose a prediction-aware testing pipeline that evaluates the underlying characteristic of a dataset while accounting for the model’s imprecision during statistical analysis. Considering the model's accuracy in our testing pipeline helps minimize the required manual labeling, enabling large-scale auditing.
The Section \ref{sec:state_of_the_art} presents the literature review about the sensitive feature extraction method and about the error-robust statistical testing. The Section \ref{sec:dataset_and_manual} introduces the datasets used and our manual annotation procedure, the Section \ref{sec:sa_classif} explains our feature extraction and classification methodology, the Section \ref{sec:testing} presents our error-aware testing protocol, then the Section \ref{sec:result} highlights our results. Section \ref{sec:conclusion} concludes with some perspectives and future work.

\section{Related Works}
\label{sec:state_of_the_art}
Assessing bias in image datasets requires careful consideration of several aspects. First, the dataset contains potentially sensitive variables. Some features must be extracted to serve as proxy estimates for these variables. Based on these features, an auditing pipeline generates reports on diversity and representativeness using selected metrics or statistical tests. The following subsections provide an overview of general concepts from the literature related to each of these aspects.

 \subsection{Choice of possible sensitive variables}
Ethnic classification refers to the classification of individuals into distinct groups based on perceived physical characteristics, such as skin color, hair texture, and facial shape. Many academic datasets separate images into at least five categories: Latino, Asian, White, Black, and Other. This classification is common in many reference datasets such as the Adult dataset \cite{misc_adult_2} and is derived from the US Census 2000 classification. In \cite{Hanna20}, the authors criticize methodologies that rely exclusively on race as a variable, arguing that this approach is overly restrictive.

An alternative to the Census 2000 classification is to use medical skin analysis criteria. In \cite{Bevan22}, the authors presented a method using the ITA (Individual Typology Angle)  algorithm \cite{thong2023,merler2019diversityfaces} to estimate skin tone in the context of classifying skin lesions and to normalize the impact of lighting variations on facial images. Similarly, the Fitzpatrick classification, introduced by \cite{Fitzpatrick_1988}, classifies individuals based on their skin's reaction to sun exposure. This classification has six classes and takes into account features such as skin color, the presence of freckles, hair and eye color, and reactions to sun exposure (precise definition and the demographic distribution are in the Appendix, Section \ref{app:sec:fitz_def}). Inspired by \cite{buolamwini_gender_2018}, we believe that the Fitzpatrick scale is well-defined as it stems from its dermatology origin. This thorough definition paired is with its popularity justify in our opinion its usage in the context of auditing.

The authors of \cite{schumann2024consensussubjectivityskintone} recommend considering ethnicity as a color shade, in particular to use the newly created Monk Skin Tone Scale \cite{monk_2023}. However, the wide range of shades makes it challenging to separate groups and, consequently, to identify bias. Once the features are selected, the next step is to automate their extraction from the image dataset.

\subsection{Model for dataset labelisation}

Depending on the size of the dataset, manual extraction may be time-consuming and challenging, prompting the use of a classifier to automatically annotate part of the dataset. CNNs are particularly well suited to images, as they can extract visually identifiable features. From a face image, CNNs can capture skin tone as a set of pixel colors or as ITA. However, this information alone omits ethnic features \cite{thong2023} such as hair texture or face shape, which can reduce the accuracy of ethnic classification.  This highlights the need for image segmentation. Therefore, the chosen architecture should identify areas that contain these features, while excluding irrelevant areas such as the background. Among existing methods \cite{Rembg}, the FairFace architecture \cite{king2015maxmarginobjectdetection} detects faces and classifies age and gender using a ResNet34 architecture \cite{he2015deepresiduallearningimage}. A variant approach in \cite{Qin_2020_PR} employs a nested U-Net architecture called $U^2$-Net. Finally, interest has brewed around understanding and guiding the CNNs by understanding how the networks treat facial characteristics \cite{zhang2017examiningcnnrepresentationsrespect}. A segmentation of the skin region can be achieved using DeepLabv3 \cite{chen2017rethinkingatrousconvolutionsemantic} with a MobileNetV3 Large Backbone model \cite{howard2019searchingmobilenetv3} pretrained on Celeb-HQ \cite{CelebAMask-HQ}. An extension proposed by  \cite{thong2023,merler2019diversityfaces} estimates ITA  values. More precisely, after smoothing the image and applying a skin mask, the authors applied a K-means clustering on the pixels values and kept the one with the highest luminosity to extract the ITA values. Finally, \cite{10776291} trained a CNN from scratch to classify skin pixel shades into 10 classes. However, none of these methods explore the impact of training dataset size which is crucial when auditing, as underlined in \cite{FaceAuditorReviewer}, or provide specific configurations for the Fitzpatrick classification. 

\subsection{Metrics and statistical tests}
 
Once features have been extracted from the dataset and transformed into variables, they are used to group individuals based on these variables. The fairness auditing process then evaluates whether certain groups are over- or under-represented in comparison to predefined parameters, which may include equal or official proportions. This parameter ensures the preservation of the so-called {\it diversity} \cite{Clemmensen2022DataRF}, such as maintaining almost equal frequencies between different groups. Consider a dataset $\mathcal{D}$  of observations composed of $p$ variables : $X^0, \dots,X^{p-1}$. Let $X^0$  be a variable that may convey bias (e.g., ethnicity or age), and  $X^j$ be a variable that may induce disparity or the bias representation (e.g., gender). We focus on the conditional distribution of $X^0$ given $X^j$, denoted as $\mathcal{L}(X^0\vert X^j)$ or when no ambiguity is possible $\{X^0\vert X^j\}$.

The first measure of fairness aims at quantifying the diversity in the dataset. For this, a diversity loss is introduced in ~\cite{Zameshina22}. Given classes $\{1, \ldots, k\}$ with target frequencies $f_i\left(\sum_{i=1}^k f_i=1\right)$, and real frequencies $f_1^{\prime}, \ldots, f_k^{\prime}$, the diversity loss $\Delta$ is defined as $\Delta:=1-\inf _{f_i>0} f_i^{\prime} / f_i$. Hence, it computes a ratio $\Delta \in [0,1]$ where a value of $\Delta$ close to 1 means that at least one group is highly under-represented. Unfortunately, this metric focuses solely on one unrepresented group. 
For discrete categorical variables, diversity can be evaluated using Conditional Shannon entropy. The Conditional Shannon entropy distribution $C(S)$ of a subset $S\subseteq X^0$ is defined as:
\[
C(S)=-\sum^k_{i=1}f_i \log f_i
\]
where $x^j_i$ is a possible modality of $X^j$ and $s_i=\frac{\vert S \vert \cap \vert X^j = x^j_i \vert}{\vert S \vert }$ is the probability to observe $S$ according $X^j = x^j_i$. Equally distributed entropy according to $X^j$ corresponds to good diversity. Both of the aforementioned metrics cannot be extended to cases where the space of conditional observations is large and are not related to a statistical test \cite{Celis16}.

The second main measure of fairness for such problems comes from a volumetric perspective comparison. Actually, Geometric diversity \cite{Clemmensen2022DataRF} provides a meaningful similarity measure for observations in multiple dimensions. Consider each data point of the dataset $x\in X$, represented by a variable vector $v_x$. The geometric diversity of a subset $S \subseteq X$ is defined as the $n$-volume of the parallelotope spanned by the $p$ variable vectors $\{v_x : x\in S\}$, where $n = \vert S \vert $ is the size of the subset. Denoting the data matrix of the subset $S$ as $\textbf{D}\in \mathbb{R}^{p \times n}$, the (squared) $n$-volume of the $n$ parallelogram embedded in $p$ dimensional space can be computed by means of the determinant of the Gramian matrix $\textbf{G} = \textbf{D}^T\textbf{D}$ (with variable vectors as columns in \textbf{D}). Thus, the geometric diversity can be measured by : 
\[
G(S)=\sqrt{\text{Det}(\textbf{D}^T\textbf{D})}
\]

The larger $G(S)$, the more diverse is $S$ in the variable space. However, Geometric Diversity cannot be applied if one aims to compare variable distributions using statistical hypothesis testing \cite{Celis2018}. The Disparate Impact (DI) is one of the most used fairness metric, defined for a binary model $\hat{Y}=f(X)$ by the ratio 
\[
DI(f,S):= \frac{\min\big(\mathbb{P}(\hat{Y} = 1 \mid S = 0), \mathbb{P}(\hat{Y} = 1 \mid S = 1)\big)}{\max\big(\mathbb{P}(\hat{Y} = 1 \mid S = 0), \mathbb{P}(\hat{Y} = 1 \mid S = 1)\big)} \]
This quantity is equal to 1 when there is probabilistic independence between the model's decision $\hat{Y}$ and the sensitive variable $S$. The smaller the DI is, the more discrimination towards the minority class exist. Hence, several norms or regulations impose that a model should have its disparate impact greater than $0.8$ as detailed in \cite{groves2024auditing} or \cite{wright2024null}. 

This metric generally used to evaluate the discrimination of a model, can be applied to evaluate the probabilistic bias of two sensitive variables. We choose to include it only in the Appendix (1) not to confuse the reader and make him think that we evaluate our model's fairness (for the instance the bias in the CNN predicting the Fitzpatrick Class), (2) homogeneity between the parity test (about one sensitive variable) where the DI is not applicable and the equal representation test (about two sensitive variables) where one might use the DI and (3) while relevant, we believe that testing a null hypothesis with multiple statistical tests is a more robust approach ; see Section \ref{app:DI} in the Appendix.


Rather than relying on high-level metrics or aggregated scores, our approach evaluates biases by directly comparing the distributions of sensitive variables across subgroups.To quantify the distance between distributions with large, high-dimensional samples, one may measure the general Wasserstein distance, given by:
\[
W_{\tilde{p}}(\mu,\nu)=\big( \inf_{\pi \in \Gamma(\mu,\nu)} \int_{M \times M} d(x,y)^{\tilde{p}} d\pi(x,y) \big)^{1/\tilde{p}} 
\]
where $\tilde{p}\geq 1$, $W_{\tilde{p}}$ is the ${\tilde{p}}^{\text{th}}$ 
Wasserstein distance, $\Gamma(\mu,\nu)$ denote all joint distributions $\pi$ that have marginals $\mu$ and $\nu$, $d()$ is the distance function between points $x$ and $y$ that matched and $M$ is a given metric space. Using the Wasserstein distance, a classical distance-based test, such as the two-sample test (i.e. variables following the same distribution), can be applied following the tests proposed in \cite{del2019central} using the limit distributions developed in \cite{del2024central} or \cite{del2019central2} and \cite{A_Statistical_Test_for_Probabilistic_Fairness}.
Other statistical tests, such as those based on averages or conditional averages, may also provide insights into the proximity of variable distributions \cite{Clemmensen2022DataRF}. Traditional statistical tests, such as Pearson’s R, the t-test, and ANOVA, are commonly used. For non-normal data distributions, non-parametric tests, such as the $\chi^2$ test, the Kolmogorov-Smirnov (KS) test, or the Central Limit Theorem (CLT) based test, serve as an alternative.

The previously mentioned metrics and tests do not account for the classification accuracy of feature extraction. Since feature extraction is performed automatically, as highlighted by \cite{adebayo2023quantifyingmitigatingimpactlabel}, who evaluated how varying levels of label error (simulated through label flipping) affected the disparity metrics, it is essential to consider the model's accuracy in the bias detection task. Permutation methods have long been used in pursuit of robustness, as demonstrated by \cite{Evaluating_Fairness_Using_Permutation_Tests}, who introduced a permutation-based fairness framework with labelled data.
Although an extensive body of work has addressed label errors in the training set, to the best of our knowledge, no specific test for bias detection, that accounts for errors in the automated extraction of variables, has been proposed .

Our contribution, therefore, is to propose a full pipeline that starts with a variable extraction step and extends to robust statistical tests designed to consider the fairness according to the accuracy of the model's annotations. The following section details our complete methodology and introduces robust statistical techniques to highlight biases in images datasets.

\section{Datasets and manual annotation}
\label{sec:dataset_and_manual}
\begin{figure}[t!]
\centering
\begin{tabular}{ccccc}
\includegraphics[width=0.15\linewidth]{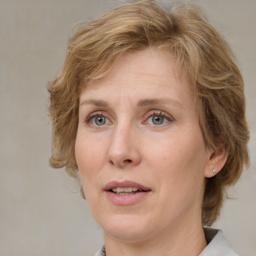} &
\includegraphics[width=0.15\linewidth]{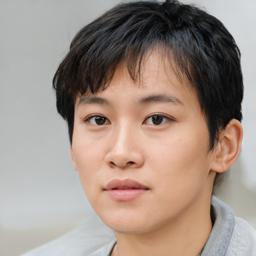} &
\includegraphics[width=0.15\linewidth]{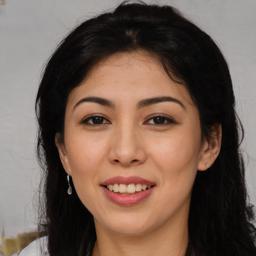} &
\includegraphics[width=0.15\linewidth]{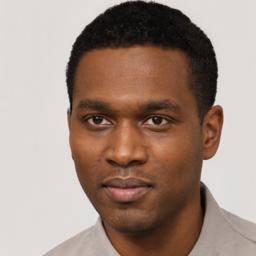} &
\includegraphics[width=0.15\linewidth]{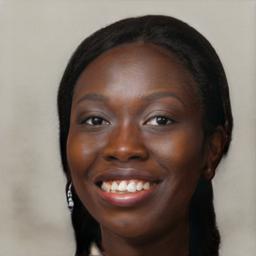} \\
\includegraphics[width=0.15\linewidth]{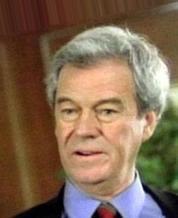} &
\includegraphics[width=0.15\linewidth]{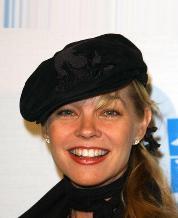} &
\includegraphics[width=0.15\linewidth]{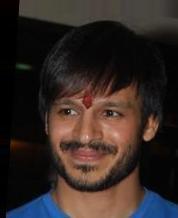} &
\includegraphics[width=0.15\linewidth]{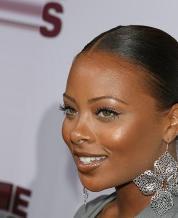} &
\includegraphics[width=0.15\linewidth]{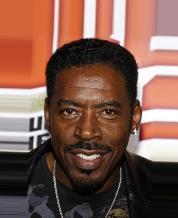} 
\end{tabular}
\caption{Fitzpatrick classification (5 class) from the left to right Phototype \rom{1}, \rom{2}, \rom{3}-\rom{4}, \rom{5}, \rom{6}. The first row is from the GAN dataset, while the second is from the CelebA dataset, both dataset are released to the community.}
\label{fig:photosFitz}
\end{figure}

\subsection{Datasets}
\label{sec:dataset}
To illustrate our pipeline, we rely on two datasets as guidelines, using them as the starting point of our process. These datasets are academic benchmark datasets of two different types. Generated Photos dataset \cite{GANDataset,boddeti2023biometriccapacitygenerativeface} is a  synthetic images dataset sourced from a commercial platform and are generated using a GAN-based model \cite{karras2019stylebasedgeneratorarchitecturegenerative}.
The dataset intentionally encompasses a broad range of demographic features, including gender and ethnicity, with the GAN's hyperparameters calibrated to represent individuals with appearances associated with diverse ethnicities. Each image has been generated with Census 2000 labeling, ensuring an almost equal proportion of Caucasian, Asian, Hispanic or Latino, and Black populations. For our work, we utilized the academic version of this dataset, which contains 10,000 generated facial images.
We also work on a well-known benchmark: The CelebA dataset \cite{liu2015faceattributes} for comparative analysis. The CelebA dataset has approximately 200,000 celebrity images sourced from the Internet, annotated with multiple facial features. From this dataset, we randomly sampled 1,500 images to assess how our test performs on a smaller dataset.

\subsection{Manual annotation}

All images were manually labeled by three non-expert individuals according to the Fitzpatrick classification. This manual annotation helps assess how well our model aligns with a fully manual annotation. However, Phototypes ~\rom{3} and ~\rom{4} are hardly distinguishable for non-experts and rarely reach full agreement. Thus, we chose to merge them. Fig. \ref{fig:photosFitz} gives some examples of our manual classification. 
As with ethnicity, a person's gender is determined by the majority vote of our three annotators. Even when considering the reflected gender, our dataset did not adequately represent the transgender or the bi-gender community, just to name a few, leading the annotators to classify the portrayal of gender to the limited view of gender binary notion that includes only men and women. In this regard, we view our work as part of initial studies towards auditing gender representation, which should further be extended in this direction in the future.

\section{Sensitive variable classification using Neural Network}
\label{sec:sa_classif}

\begin{figure}[t!]
    \centering
    \includegraphics[width=1\linewidth]{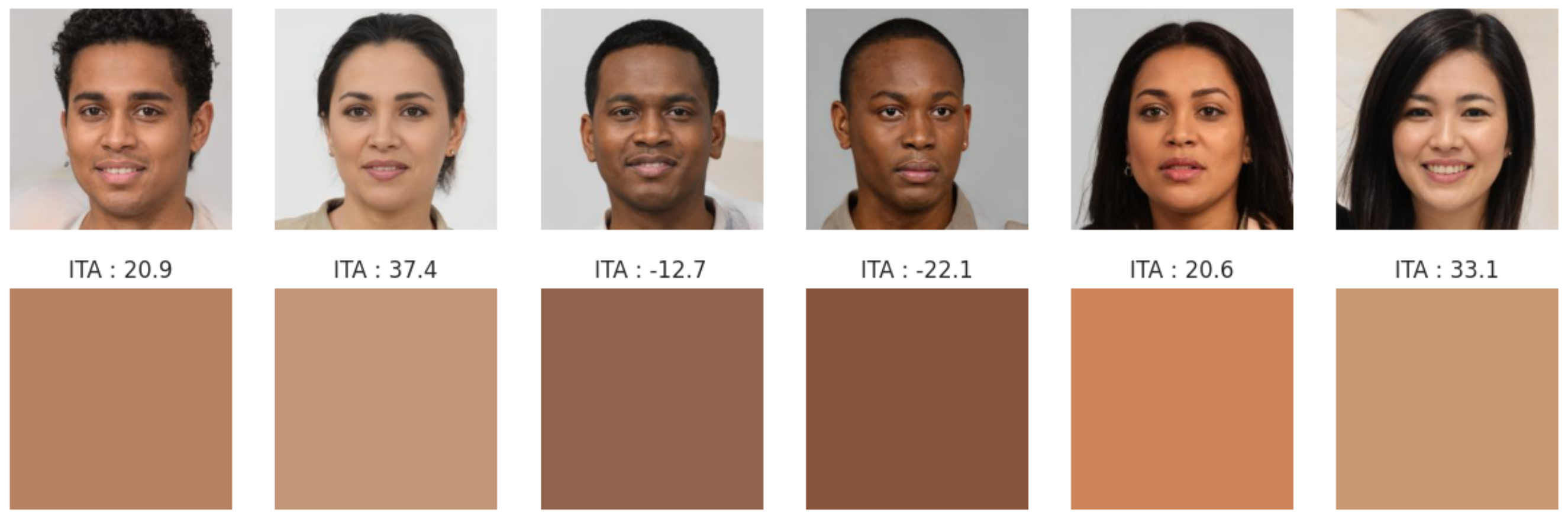}
    \caption{Skin color extracted and Individual Typology Angle (ITA) of the GAN dataset}
    \label{fig:ITAexampleGAN}
\end{figure}

\begin{figure}[t!]
    \centering
    \begin{subfigure}[t]{0.49\textwidth}
        \centering
        \includegraphics[height=1.6in]{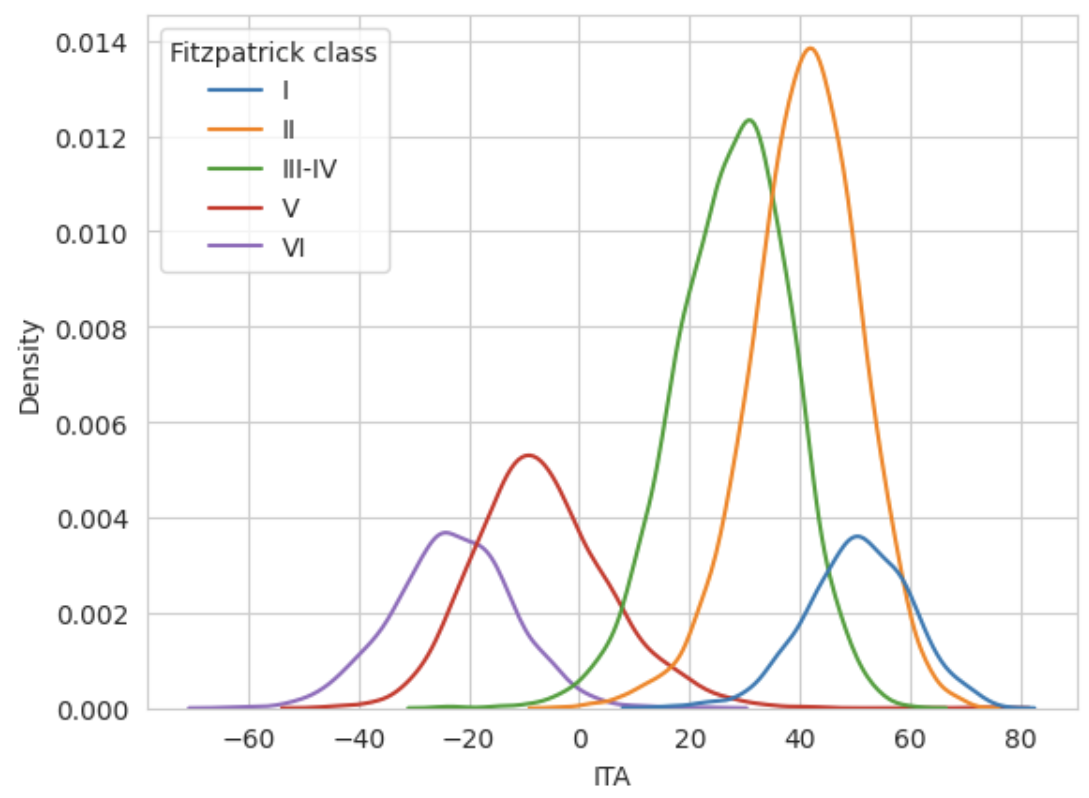}
        \caption{GAN dataset}
        \label{fig:repartitionITAFitzpatrickGAN}
    \end{subfigure}
    \hfill
    \begin{subfigure}[t]{0.49\textwidth}
        \centering
        \includegraphics[height=1.6in]{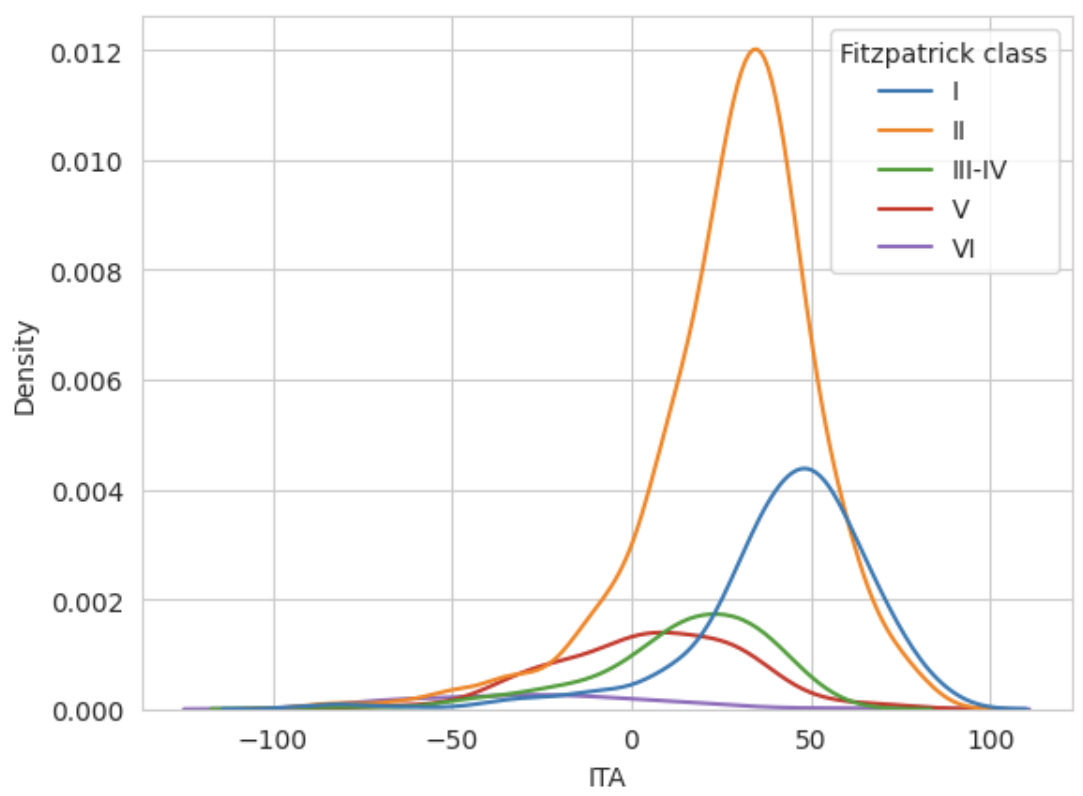}
        \caption{Subset of the CelebA dataset}
        \label{fig:repartitionITAFitzpatrickCeleba}
    \end{subfigure}
    \caption{Probability density of ITA given the Fitzpatrick class}
    \label{fig:repartitionITAFitzpatrick}
\end{figure}

Manual auditing is not cost-effective for high- or medium-level auditing of large-scale datasets. This underscores the need for neural networks to predict sensitive variables accurately. Numerous manually labelled image datasets exist for binary gender classification - although we regret the lack of datasets with more diverse gender representations - facilitating the use of highly effective pre-trained networks for gender estimation.

\subsection{Individual Typology Angle (ITA) estimation and link with Fitzpatrick}
To improve the accuracy of the Fitzpatrick classification, we add an Individual Typology Angle (ITA) estimation step to our model, which can be seen as an enrichment of pixel information.  ITA values are computationnaly derived from skin regions (isolated by pre-trained DeepLabv3). The estimate of the ITA value is based on 2 colorimetric parameters: the luminance $L*$ and the yellow/blue component $b*$. The ITA is defined as follows:
\begin{equation}
    \text{ITA} = \text{arctan}\bigg(\frac{L* - 50}{b*}\bigg) \times \frac{180}{\pi}
\end{equation}
where a perceptual lightness at value 50 corresponds to a maximum chroma. We extracted the mean and standard deviation of the ITA values and of other colometric parameters. Fig. \ref{fig:ITAexampleGAN} presents examples of extracted ITA.

Fig. \ref{fig:repartitionITAFitzpatrick} gives the ITA distribution according the Fitzpatrick class and confirm the clear correlation between the Fitzpatrick classes and ITA values. Higher Fitzpatrick class numbers correspond to lower mean ITA values. However, as a single ITA score can be assigned to several Fitzpatrick classes, there is no one-to-one correspondence between the two.  We further research the difference between the ITA and the Fitzpatrick class in the Appendix, Section \ref{app:section:ITA_Fitz}.

\subsection{Gender, age and Fitzpatrick scale classification} \label{sec:FPforecast}

\begin{figure}[t!]
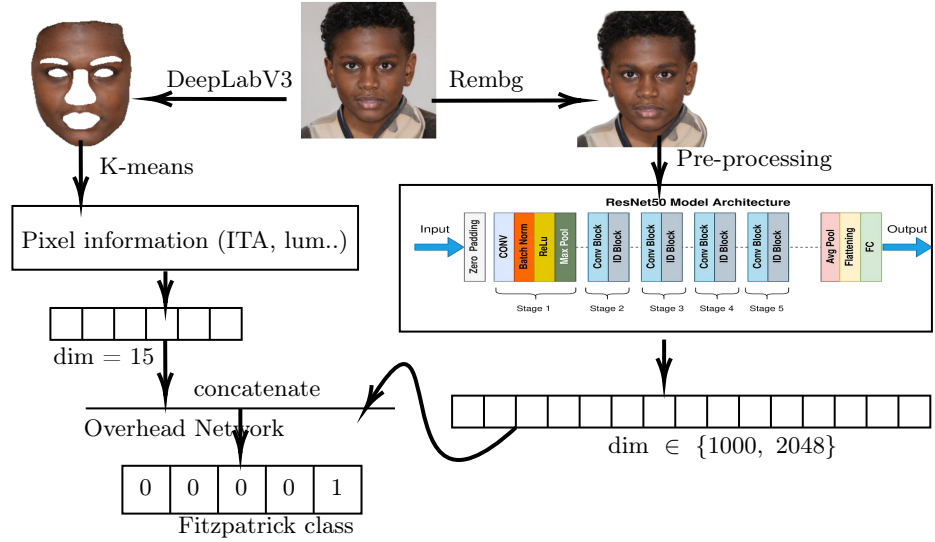

    \centering
    \include{Diagram/Architecture_smaller}
    \caption{Pipeline for our Fitzpatrick classification}
    \label{fig:Architecture}
\end{figure}

We used the FairFace \cite{karkkainenfairface} method to classify gender and age, as its architecture provides the best results for these tasks. Despite certain limitations —such as detecting undesirable background faces- it achieved the best accuracy (see in the Appendix, Section \ref{app:section:FairFace}). Since our classification task relies primarily on facial features, especially skin, we studied the effect of applying masks to images before training our fine-tuned CNN. We tested three approaches (see in the Appendix, Fig. \ref{fig:mask}): (1) using the original images, (2) removing the background, and (3) isolating only the segmented skin region. The extracted ITA and skin-related information are incorporated as additional features in the latent layer of the neural network architecture.

In most of our experiments, we added a custom classification head to ResNet-50 or ResNet-101 embeddings \cite{he2015deepresiduallearningimage} and fine-tuned these models on our labeled dataset, eliminating the need for full CNN training.  The impact of training set size is analyzed in Section \ref{sec:result}. We use two feature extraction configurations: (1) the final dense layer's output (1,000-dimensional) and (2) the preceding layer's output (2,048-dimensional). Fig. \ref{fig:Architecture} present an overview of our classification pipeline. More details on the architecture, optimizer, early stopping, compute time, and transfer learning method are provided in the Appendix, Section \ref{app:method:classif}. We created a Neural Network architecture to accurately predict the Fitzpatrick class with as few manually labeled image as possible, however, this part was not mandatory to our auditing framework thanks to the following section, which explains how we calibrate our testing pipeline given a model's accuracy.

\section{Uncertainty aware statistical test}
\label{sec:testing}

\begin{figure}[t!]
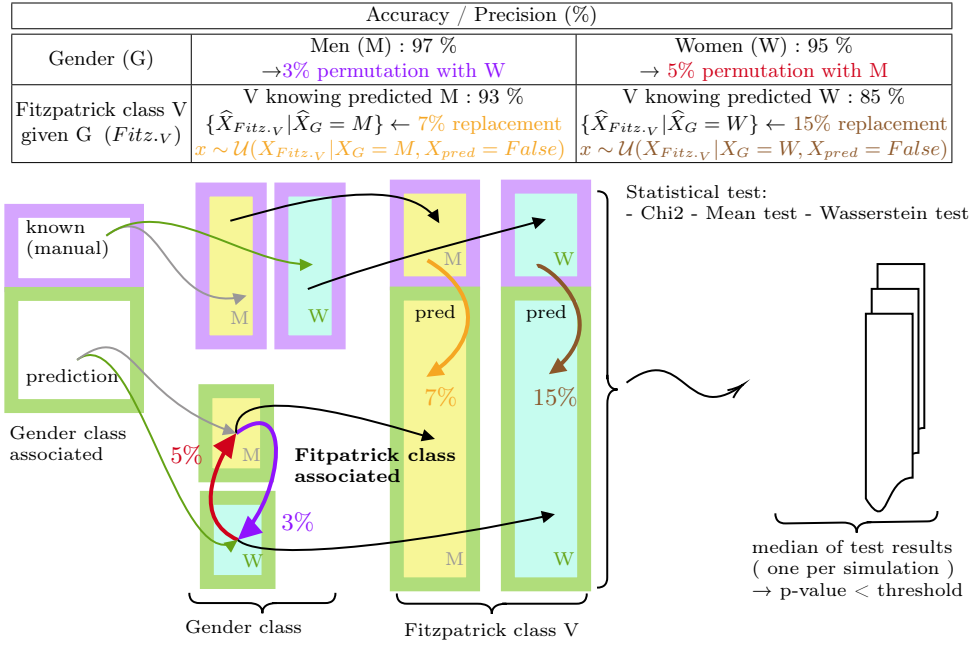

    \centering
    \include{Diagram/test_regular}
    \caption{Diagram explaining our error-aware testing pipeline in an equal representation test of the Fitzpatrick class \rom{5} conditioned by the gender.}
    \label{tst_diagram}
\end{figure}

\paragraph{Statistical test used}

We find rejection based on variable distributions more meaningful, hence, $\mathcal{H}_0$ assumes that both groups are drawn from the same underlying distribution. In our proposed methodology, we use a modified version of well-known statistical tests to compare two distributions, including the  \(\chi^2\) test, the CLT-based mean test and the Wasserstein-based test. Note that categorical multimodal variables are treated as binary variables in a one-vs-all approach, which can be considered a limitation to our work. Two tests are presented here: the parity test and the equal representation test.

\paragraph{Parity test (one sensitive variable)}

To audit the bias according to a tested variable $X^0$ (gender, age or Fitzpatrick), it is necessary to compare the observed distribution of $X^0$ with its expected  distribution. When $X^0$ is \textit{gender}, this process involves comparing $X^0$ observed values with a Bernoulli distribution of $p = \frac{1}{2}$ (i.e. testing $H_0:X^0 \sim \mathcal{B}(p)$).
While the assumption of a uniform distribution for the binary gender may appear reasonable, it is important to recognize the limitations of such an assumption when considering age groups or the Fitzpatrick class. To this end, a chosen parameter reflecting a real mondial distribution was utilized for comparison (see in the Appendix, in Section \ref{app:section:Result_audit}, Table \ref{tab:age_selon_genre} and Table \ref{tab:Fitzpatrick_selon_genre} to observe the recorded parameters). Hence, we test, respectively for when $X^0$ is the age ($H_0:X^0 \sim RealDistr(Age)$) and for when $X_0$ is the Fitzpatrick class ($H_0: X^0 \sim RealDistr(Fitzpatrick)$).

\paragraph{Equal representation (two sensitive variables)}

We test whether the distribution of a variable $X^0$ (a Fitzpatrick skin type or age interval) differs significantly given another variable $X^j$ (gender 'men' or 'women'). 
Let's consider $x^0_{i'} \in 1,\cdots,K'$ the $K'$ modalities of $X^0$ and $x^j_i \in 1,\cdots,K $ the $K$ modalities of $X^j$. To perform this analysis, we first partition the dataset based on $X^j$ (one-versus-all according to $x_i^j$) and then compare the distribution of $X^0$ across the two resulting partitions of $X^j$.  
We define the variable $W^0 \in \{0,1\}^{K'}$
such as $W^0 = (W^0_1,\cdots , W^0_{K'})$ and the $W^0_{i'}$ are defined as followed:
\begin{equation}
  \forall i' \in 1,\cdots , K' \hspace{0.5cm} W^0_{i'}:=\begin{cases}
    1 & \text{if $X^0 = x^0_{i'}$}\\
    0 & \text{otherwise}.
  \end{cases}
\end{equation}
a condition notation of $W^0_{i'}$  on a subspace $S$ according $X^j$ is given by : 
\begin{align}
  \forall i \in 1,\cdots , K \hspace{0.5cm} &W^{0,j}_{i',i}:=\begin{cases}
    1 & \text{if $X^0 = x^0_{i'}$ in  $S\in \{X^j=x^j_{i}\}$}\\
    0 & \text{if $X^0 \neq x^0_{i'}$ in  $S\in \{X^j=x^j_{i}\}$}\\
  \end{cases}\\
 &W^{0,j}_{i',\ib}:=\begin{cases}
    1 & \text{if $X^0 = x^0_{i'}$  in $S \in \{X^j\neq x^j_{i}\}$}\\
    0 &  \text{if $X^0 \neq x^0_{i'}$  in  $S \in \{X^j\neq x^j_{i}\}$}\\
  \end{cases}
\end{align}


We test the following assumption on the distributions, 
$H_0 : W^{0,j}_{{i'},i} \sim W^{0,j}_{{i'},\ib}$
.

\paragraph{Uncertainty aware}
To ensure reliability, the auditing process must be robust to variations in model annotation accuracy. Consequently, the test must be robust to prediction errors and minimize false negatives for the null hypothesis, $ \mathcal{H}_0 $. 
As permutation tests, we randomly invert the automatic annotation modality of some predicted variables while keeping manual annotations unchanged. This procedure serves to reduce the discrepancy between distributions and minimize false negatives in test decisions. The model prediction is denoted by $\widehat{W}^0_i$ , and the true value by $W^0_i$. These tests are modified according to the following methodology:

\begin{itemize}
    \item Parity test: \begin{enumerate}
        \item We calculate the model's accuracy $A^{W^0_{i'}} := \mathbb{P}[\widehat{W}^0_{i'} = W^0_{i'}]$ for each of the estimated variables $\widehat{W}^0_{i'}$. 
        \item We randomly replace $100 \times (1-A^{W^0_{i'}})\%$ of $\widehat{W}^0_{i'}$ by values simulated according to the expected parameter (for example $\mathcal{B}(\frac{1}{2})$ for gender).
        \end{enumerate}
        
    \item Representation test: \begin{enumerate}
        \item  We calculate the prediction's precision $P^{W^j_{i}} := \mathbb{P}[ W^j_i = 1| \widehat{W}^j_i = 1] $  for each of the variables $\widehat{W}^j_{i}$.
        \item We randomly permute $100 \times (1-P^{W^j_{i}})\%$ between $\widehat{W}^j_{i}$ and $\widehat{W}^j_{\ib}$ values (e.g., transforming predicted women into predicted men and vice versa)
        \item We compute $A^{W^0_{i'},W^j_i=1} := \mathbb{P}[\widehat{W}^0_{i'} = W^0_{i'}|W^j_i = 1]$ which represents the classification model's accuracy for the modality $x^0_{i'}$ conditioned on $x^j_i$.
        \item We randomly replace $100 \times (1-A^{W^0_{i'},W^j_i=1})\%$ of the automatically annotated variables $\widehat{W}^{0,j}_{i',i}$ and $\widehat{W}^{0,j}_{i',\ib}$ respectively with manually annotated variables $W^{0,j}_{i',i}$ and $W^{0,j}_{i',\ib}$.
    \end{enumerate}
\end{itemize}


Note that the accuracies and precisions above are calculated on the validation set. We conduct multiple statistical tests across several simulations and aggregate the results by taking the median p-value of all simulations. This framework is illustrated in Fig. \ref{tst_diagram}.  

\section{Results}
\label{sec:result}

We report results on the two datasets described in Section \ref{sec:dataset}.  Our auditing process consists of: (1) assessing whether there is a significant difference in the observed proportions across three sensitive attributes—gender, age, and Fitzpatrick classification; and (2) evaluating whether significant differences exist in the observed proportions of age and Fitzpatrick classification, \textbf{conditioned on gender}.

\subsection{Ablation study of the Fitzpatrick classification}
\begin{figure}[t!]
    \centering
    \begin{minipage}{0.49\textwidth}
        \centering
        \begin{subfigure}[t]{0.49\textwidth}
            \centering
            \includegraphics[width=0.99\linewidth]{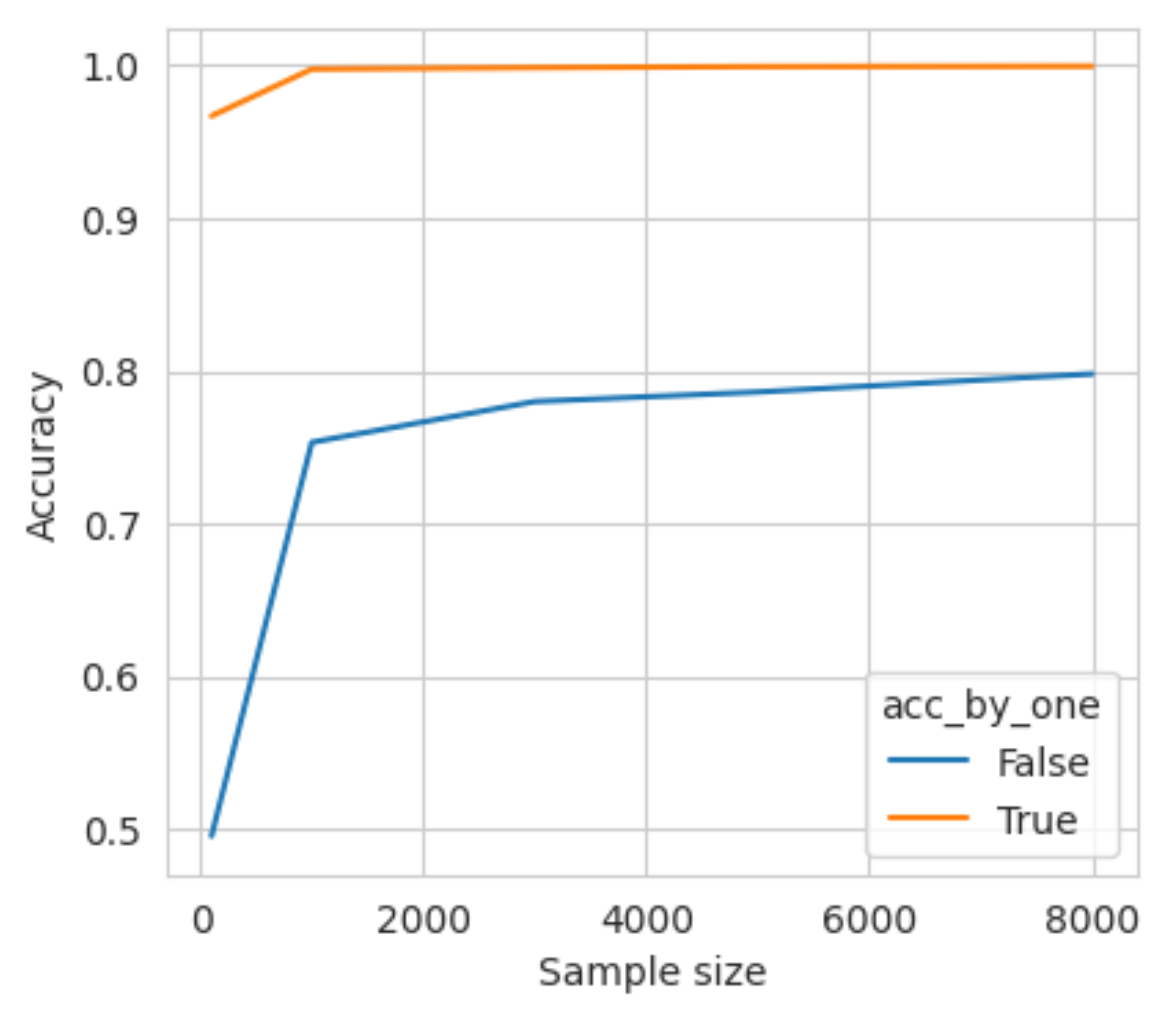}
            \caption{GAN dataset}
            \label{fig:sample_size_GAN}
        \end{subfigure}
        \hfill
        \begin{subfigure}[t]{0.49\textwidth}
            \centering
            \includegraphics[width=0.99\linewidth]{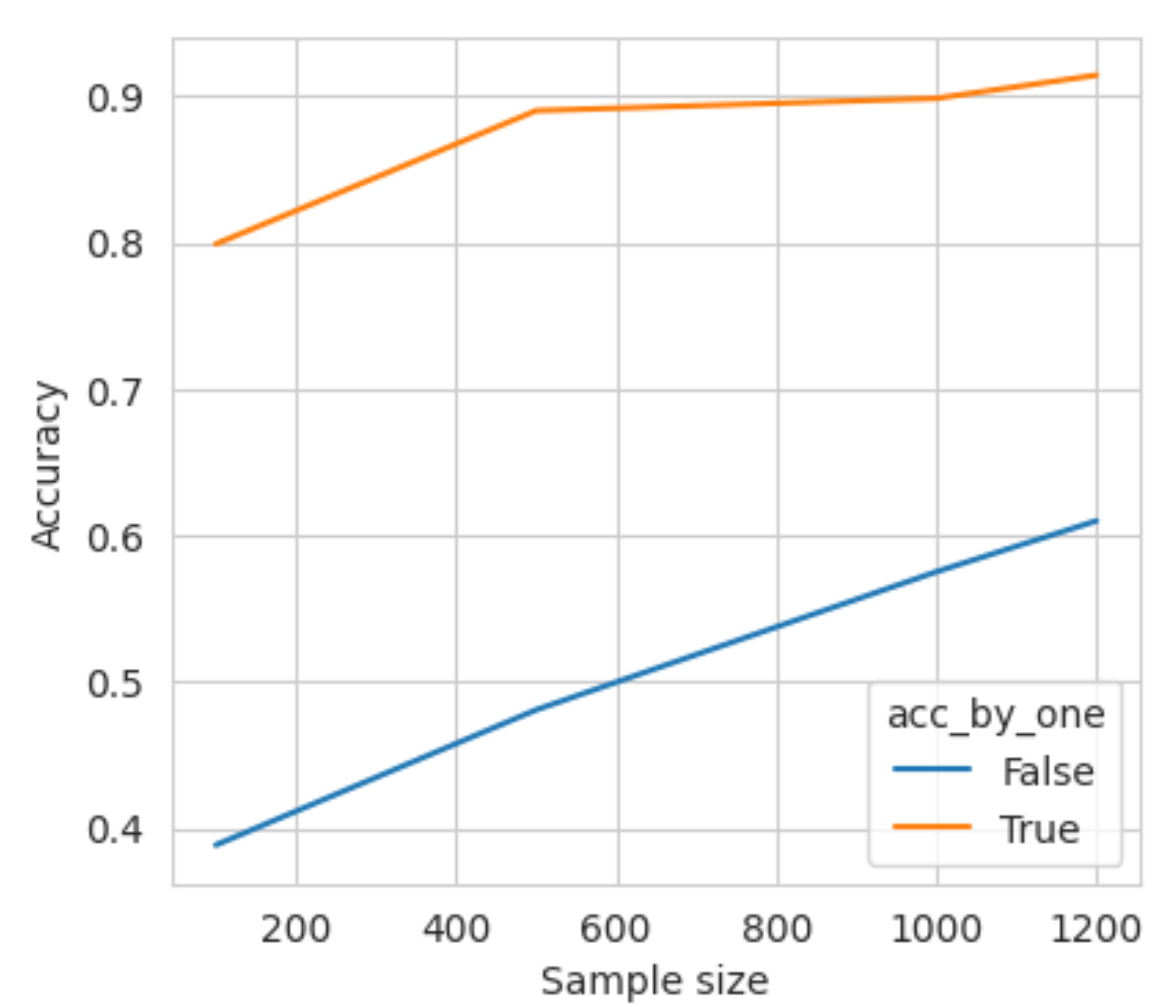}
            \caption{CelebA dataset}
            \label{fig:sample_size_CelebA}
        \end{subfigure}
        \caption{Learning sample size impact on the Network accuracy (Fitzpatrick classification). The by-one-accuracy includes predictions for the true class as well as the directly adjacent classes.
        }
        \label{fig:sample_size}
    \end{minipage}
    \hfill
    \begin{minipage}{0.49\textwidth}
        \centering
        \scalebox{1}{\begin{tabular}{c|cc}
            \toprule
            \textbf{Hyperparameter} & \textbf{ GAN } & \textbf{ CelebA } \\
            \midrule
            Skin information (ITA..)  & +0.6  & +0.2  \\
            Overhead network choice   & +3.7  & +4.2  \\
            Skin mask                 & -1.7  & +1.2  \\
            Removing background       & +0.1  & +1.8  \\
            Latent space size (2048)  & +1.9  & +2.7  \\
            \bottomrule
        \end{tabular}}
        \captionof{table}{Impact of hyperparameters and architecture for neural network designed to classify the Fitzpatrick scale on the GAN and CelebA datasets.}
        \label{tab:neural_network_ablation_study}
    \end{minipage}
\end{figure}

As shown in Fig.\ref{fig:sample_size}, the size of the manually annotated training dataset has a significant influence on model accuracy. Using the GAN dataset, our model achieves a 76\% correct Fitzpatrick classification prediction rate on the test set, with at least 12.5\% manual annotation. However, with CelebA, the accuracy of the model does not exceed 65\%, despite hyperparameter tuning. It highlights the necessity of accounting for model errors in the tests to avoid the need for a full retraining of the CNN. Age and gender are provided by a trained FairFace model, so there is no training step, and the model achieved respectively a 97.17\% and a 94.46\% accuracy on \textit{CelebA} and \textit{GAN} for the gender classification. Without age labels, to verify the consistency of FairFace's age prediction, we compared its prediction with another network prediction and obtained a 72.77\% classification similarity for the GAN dataset (More details in the Appendix, see Table \ref{tab:confusion_matrix_predictions_age}).
Table  \ref{tab:neural_network_ablation_study} provides an ablation study examining the impact of each hyperparameter in the CNN architecture for Fitzpatrick classification. A high-dimensional latent space, the incorporation of ITA into the model, and the removal of image backgrounds significantly enhance the learning process. However, the use of a skin mask has a negative effect on the results.  This can be explained by the removal of features such as hair color, which are essential for Fitzpatrick classification. The addition of ITA showed a notable improvement for small training datasets with an embedding size of 1000 dimensions. We believe that skin-related features are sufficiently captured for the embedding of size 2048 but lost during dimensionality reduction (embedding of size 1000).  

\subsection{Ablation study of the error-aware method}

\begin{table}[t!]
    \centering
    \caption{Statistical test on the parity of Gender ($\mathcal{H}_0$ : $p=0.5$) and the real distribution for the Fitzpatrick class and Age ($\mathcal{H}_0$ : marginal and global distributions are equivalent). The test used were the Wasserstein test, the Mean test and the \(\chi^2\) test. {$\checkmark$}, {$\checkmark_{2/3}$} and ${\color{RedOrange}\times}$ respectively means that 0, 1 or at least 2 tests rejected $\mathcal{H}_0$.  Colored cell means that the test result are different because of the error-aware protocol:  \crule[LimeGreen]{0.3cm}{0.3cm} (with $-$ within) means that, thanks to the error-aware method, less test rejected $\mathcal{H}_0$.}
    \label{tab:parity_tests}
    \scalebox{0.9}{
    \begin{minipage}{0.58\textwidth}
        \centering
        \subcaption{GAN}
          \label{tab:parity_tests_GAN}
        \setlength{\tabcolsep}{4pt}
        \begin{tabular}{|l|c|c|c|c|c|}
            \hline
            \multirow{2}{*}{\textbf{\shortstack{Sensitive\\Variable}}} & 
            \multicolumn{5}{c|}{\textbf{Sample Size}} \\ 
            \cline{2-6}
            & \textbf{100} & \textbf{500} & \textbf{1000} & \textbf{3000} & \textbf{8000} \\ 
            \hline
            \textbf{Gender} & ${\color{RedOrange}\times}$ & ${\color{RedOrange}\times}$ & ${\color{RedOrange}\times}$ & ${\color{RedOrange}\times}$ & ${\color{RedOrange}\times}$ \\  
            \hline
            \textbf{Age} & & & & & \\ 
            \quad 0-2 & \cellcolor{LimeGreen} {$\checkmark$} \makebox[0pt][c]{\smash[b]{\rule{0pt}{1.5ex}{\color{white}$-$}}} & \cellcolor{LimeGreen} {$\checkmark$} \makebox[0pt][c]{\smash[b]{\rule{0pt}{1.5ex}{\color{white}$-$}}} & \cellcolor{LimeGreen} {$\checkmark$} \makebox[0pt][c]{\smash[b]{\rule{0pt}{1.5ex}{\color{white}$-$}}} & \cellcolor{LimeGreen} {$\checkmark$} \makebox[0pt][c]{\smash[b]{\rule{0pt}{1.5ex}{\color{white}$-$}}} & \cellcolor{LimeGreen} {$\checkmark$} \makebox[0pt][c]{\smash[b]{\rule{0pt}{1.5ex}{\color{white}$-$}}} \\ 
            \quad 3-9 & ${\color{RedOrange}\times}$ & ${\color{RedOrange}\times}$ & ${\color{RedOrange}\times}$ & ${\color{RedOrange}\times}$ & ${\color{RedOrange}\times}$ \\ 
            \quad 10-19 & ${\color{RedOrange}\times}$ & ${\color{RedOrange}\times}$ & ${\color{RedOrange}\times}$ & ${\color{RedOrange}\times}$ & ${\color{RedOrange}\times}$ \\ 
            \quad 20-29 & ${\color{RedOrange}\times}$ & ${\color{RedOrange}\times}$ & ${\color{RedOrange}\times}$ & ${\color{RedOrange}\times}$ & ${\color{RedOrange}\times}$ \\ 
            \quad 30-39 & ${\color{RedOrange}\times}$ & ${\color{RedOrange}\times}$ & ${\color{RedOrange}\times}$ & ${\color{RedOrange}\times}$ & ${\color{RedOrange}\times}$ \\ 
            \quad 40-49 & ${\color{RedOrange}\times}$ & ${\color{RedOrange}\times}$ & ${\color{RedOrange}\times}$ & ${\color{RedOrange}\times}$ & ${\color{RedOrange}\times}$ \\ 
            \quad 50-59 & ${\color{RedOrange}\times}$ & ${\color{RedOrange}\times}$ & ${\color{RedOrange}\times}$ & ${\color{RedOrange}\times}$ & ${\color{RedOrange}\times}$ \\ 
            \quad 60-69 & ${\color{RedOrange}\times}$ & ${\color{RedOrange}\times}$ & ${\color{RedOrange}\times}$ & ${\color{RedOrange}\times}$ & ${\color{RedOrange}\times}$ \\ 
            \quad 70+ & ${\color{RedOrange}\times}$ & ${\color{RedOrange}\times}$ & ${\color{RedOrange}\times}$ & ${\color{RedOrange}\times}$ & ${\color{RedOrange}\times}$ \\ 
            \hline
            \textbf{Fitzp. class} & & & & & \\ 
            \quad ~\rom{1} & ${\color{RedOrange}\times}$ & ${\color{RedOrange}\times}$ & ${\color{RedOrange}\times}$ & ${\color{RedOrange}\times}$ & ${\color{RedOrange}\times}$ \\ 
            \quad ~\rom{2} & ${\color{RedOrange}\times}$ & ${\color{RedOrange}\times}$ & ${\color{RedOrange}\times}$ & ${\color{RedOrange}\times}$ & ${\color{RedOrange}\times}$ \\ 
            \quad ~\rom{3}-~\rom{4} & ${\color{RedOrange}\times}$ & ${\color{RedOrange}\times}$ & ${\color{RedOrange}\times}$ & ${\color{RedOrange}\times}$ & ${\color{RedOrange}\times}$ \\ 
            \quad ~\rom{5}& ${\color{RedOrange}\times}$ & ${\color{RedOrange}\times}$ & ${\color{RedOrange}\times}$ & ${\color{RedOrange}\times}$ & \cellcolor{LimeGreen} {$\checkmark$} \makebox[0pt][c]{\smash[b]{\rule{0pt}{1.5ex}{\color{white}$-$}}} \\ 
            \quad ~\rom{6}& ${\color{RedOrange}\times}$ & ${\color{RedOrange}\times}$ & ${\color{RedOrange}\times}$ & ${\color{RedOrange}\times}$ & ${\color{RedOrange}\times}$ \\ 
            \hline
        \end{tabular}
    \end{minipage}
    \hfill
    \begin{minipage}{0.38\textwidth}
        \centering
        \subcaption{CelebA}
          \label{tab:parity_tests_CelebA}
        \setlength{\tabcolsep}{6pt}
        \begin{tabular}{|c|c|c|c|}
            \hline
            \multicolumn{4}{|c|}{\textbf{Sample Size}} \\ 
            \cline{1-4}
            \textbf{100} & \textbf{500} & \textbf{1000} & \textbf{1200} \\ 
            \hline
            ${\color{RedOrange}\times}$ & ${\color{RedOrange}\times}$ & ${\color{RedOrange}\times}$ & ${\color{RedOrange}\times}$ \\ 
            \hline
            & & & \\ 
            \cellcolor{LimeGreen} {$\checkmark$} \makebox[0pt][c]{\smash[b]{\rule{0pt}{1.5ex}{\color{white}$-$}}} & \cellcolor{LimeGreen}{$\checkmark$} \makebox[0pt][c]{\smash[b]{\rule{0pt}{1.5ex}{\color{white}$-$}}} & \cellcolor{LimeGreen} {$\checkmark$} \makebox[0pt][c]{\smash[b]{\rule{0pt}{1.5ex}{\color{white}$-$}}} & \cellcolor{LimeGreen} {$\checkmark$} \makebox[0pt][c]{\smash[b]{\rule{0pt}{1.5ex}{\color{white}$-$}}} \\ 
            ${\color{RedOrange}\times}$ & ${\color{RedOrange}\times}$ & ${\color{RedOrange}\times}$ & ${\color{RedOrange}\times}$ \\ 
            ${\color{RedOrange}\times}$ & ${\color{RedOrange}\times}$ & ${\color{RedOrange}\times}$ & ${\color{RedOrange}\times}$ \\ 
            ${\color{RedOrange}\times}$ & ${\color{RedOrange}\times}$ & ${\color{RedOrange}\times}$ & ${\color{RedOrange}\times}$ \\ 
            ${\color{RedOrange}\times}$ & ${\color{RedOrange}\times}$ & ${\color{RedOrange}\times}$ & ${\color{RedOrange}\times}$ \\ 
            \cellcolor{LimeGreen} {$\checkmark$} \makebox[0pt][c]{\smash[b]{\rule{0pt}{1.5ex}{\color{white}$-$}}} & \cellcolor{LimeGreen} {$\checkmark$} \makebox[0pt][c]{\smash[b]{\rule{0pt}{1.5ex}{\color{white}$-$}}} & \cellcolor{LimeGreen} {$\checkmark$} \makebox[0pt][c]{\smash[b]{\rule{0pt}{1.5ex}{\color{white}$-$}}} & \cellcolor{LimeGreen} {$\checkmark$} \makebox[0pt][c]{\smash[b]{\rule{0pt}{1.5ex}{\color{white}$-$}}} \\ 
            \cellcolor{LimeGreen} {$\checkmark$} \makebox[0pt][c]{\smash[b]{\rule{0pt}{1.5ex}{\color{white}$-$}}} & \cellcolor{LimeGreen} {$\checkmark$} \makebox[0pt][c]{\smash[b]{\rule{0pt}{1.5ex}{\color{white}$-$}}} & \cellcolor{LimeGreen} {$\checkmark$} \makebox[0pt][c]{\smash[b]{\rule{0pt}{1.5ex}{\color{white}$-$}}} & \cellcolor{LimeGreen} {$\checkmark$} \makebox[0pt][c]{\smash[b]{\rule{0pt}{1.5ex}{\color{white}$-$}}} \\ 
            ${\color{RedOrange}\times}$ & ${\color{RedOrange}\times}$ & ${\color{RedOrange}\times}$ & ${\color{RedOrange}\times}$ \\ 
            ${\color{RedOrange}\times}$ & ${\color{RedOrange}\times}$ & ${\color{RedOrange}\times}$ & ${\color{RedOrange}\times}$ \\ 
            \hline
            & & & \\ 
            ${\color{RedOrange}\times}$ & ${\color{RedOrange}\times}$ & ${\color{RedOrange}\times}$ & ${\color{RedOrange}\times}$ \\ 
            ${\color{RedOrange}\times}$ & ${\color{RedOrange}\times}$ & ${\color{RedOrange}\times}$ & ${\color{RedOrange}\times}$ \\ 
            ${\color{RedOrange}\times}$ & ${\color{RedOrange}\times}$ & ${\color{RedOrange}\times}$ & ${\color{RedOrange}\times}$ \\ 
            
            ${\color{RedOrange}\times}$ & ${\color{RedOrange}\times}$ & ${\color{RedOrange}\times}$ & \cellcolor{LimeGreen} {$\checkmark$} \makebox[0pt][c]{\smash[b]{\rule{0pt}{1.5ex}{\color{white}$-$}}} \\ 
            ${\color{RedOrange}\times}$ & ${\color{RedOrange}\times}$ & ${\color{RedOrange}\times}$ & \cellcolor{LimeGreen} {$\checkmark_{2/3}$} \makebox[0pt][c]{\smash[b]{\rule{0pt}{1.5ex}{\color{white}$-$}}} \\ 
            \hline
        \end{tabular}
    \end{minipage}}
\end{table}

To study the impact of the error-aware approach, we evaluated the statistical tests with and without taking into account the imprecision of the neural network's predictions. The colored cell on Table \ref{tab:parity_tests} and Table \ref{tab:result_eoo} show the effect of adding the uncertainty aware corrections to the statistical tests. For the parity test (1), for over the 135 tests, the aggregation of test accepted $\mathcal{H}_0$ 20 times with the error-aware approach, and only six times without it. For the equal representation test (2), the error-aware method  affected the result of 88 of the 126 aggregations of tests: for 84 out of the previously mentioned 88, the uncertainty aware corrections made the audit result more tolerant. we provide in the Appendix the Table \ref{tab:parity_tests} and Table \ref{tab:result_eoo} without corrections (Table \ref{app:tab:parity_tests} and Table \ref{app:tab:result_eoo}).

\subsection{Sample size impact on statistical test results}

Here, we assess whether both test methodologies produce the same conclusions as those obtained from the fully annotated dataset, which serves as the ground truth without annotation errors, given different amounts of manually annotated data. The parity test (1), for the \textit{GAN} dataset resulted in only four false rejections out of the 75 tested hypotheses, with the error associated with Fitzpatrick category \rom{5} (Table \ref{tab:parity_tests_GAN}). 
For the \textit{CelebA} dataset, we observed two modalities with false rejection out of the 15 tested (see Table \ref{tab:parity_tests_CelebA}). In both datasets, our parity test demonstrates robustness to the \textit{sample size} effect. The equal representation test (2) is more sensitive to sample size effect. For the \textit{GAN} dataset, it produced eleven false rejections out of 70 tests (Table \ref{tab:result_eooGAN}). 
It seems that sample size$\geq1000$ is enough to get stabilized results. For the \textit{CelebA} dataset, Our equal representation test (Table \ref{tab:result_eoo}) produced three false negative and three false positives among 56 tests.

\subsection{Auditing result}

\begin{table}[t!]
    \centering
    \caption{Equal representation statistical test on the  for the Fitzpatrick class and the Age, with respect to each reflected closest binary gender subgroup. $\mathcal{H}_0$: The Fitzpatrick or Age distribution of the reflected men subgroup is the same as the reflected women subgroup. The tests used were the Wasserstein test, the Mean test, and the \(\chi^2\) test. {\color{black}$\checkmark$}, {\color{black}$\checkmark_{2/3}$} and ${\color{RedOrange}\times}$ respectively means that 0, 1 or at least 2 tests rejected $\mathcal{H}_0$. 
    Colored cell means that the test result are different because of the error-aware protocol: \crule[LimeGreen]{0.3cm}{0.3cm} (with $-$ within) and respectively \crule[pink]{0.3cm}{0.3cm} (with $+$ within) mean that, because of the error-aware method, less test or respectively more test rejected $\mathcal{H}_0$}
    \label{tab:result_eoo}
    \scalebox{0.9}{\begin{subtable}{0.60\textwidth}
        \centering
        \caption{GAN}
            \label{tab:result_eooGAN}
        \setlength{\tabcolsep}{4pt} 
        \begin{tabular}{|l|c|c|c|c|c|}
            \hline
            \multirow{2}{*}{\textbf{\shortstack{Sensitive\\Variable}}} & 
            \multicolumn{5}{c|}{\textbf{Sample Size}} \\ 
            \cline{2-6}
            & \textbf{100} & \textbf{500} & \textbf{1000} & \textbf{3000} & \textbf{8000} \\ 
            \hline
            \textbf{Age} & & & & & \\ 
            \quad 0-2 & \cellcolor{LimeGreen}{\color{black}$\checkmark$} \makebox[0pt][c]{\smash[b]{\rule{0pt}{1.5ex}{\color{white}$-$}}}& 
            \cellcolor{LimeGreen}{\color{black}$\checkmark$} \makebox[0pt][c]{\smash[b]{\rule{0pt}{1.5ex}{\color{white}$-$}}}& 
            \cellcolor{LimeGreen}{\color{black}$\checkmark$} \makebox[0pt][c]{\smash[b]{\rule{0pt}{1.5ex}{\color{white}$-$}}}& 
            \cellcolor{LimeGreen}{\color{black}$\checkmark$} \makebox[0pt][c]{\smash[b]{\rule{0pt}{1.5ex}{\color{white}$-$}}}& 
            \cellcolor{LimeGreen}{\color{black}$\checkmark$} \makebox[0pt][c]{\smash[b]{\rule{0pt}{1.5ex}{\color{white}$-$}}}\\ 
            
            \quad 3-9 & \cellcolor{LimeGreen}{\color{black}$\checkmark$} \makebox[0pt][c]{\smash[b]{\rule{0pt}{1.5ex}{\color{white}$-$}}}& 
            \cellcolor{LimeGreen}{\color{black}$\checkmark$} \makebox[0pt][c]{\smash[b]{\rule{0pt}{1.5ex}{\color{white}$-$}}}& 
            \cellcolor{pink}${\color{RedOrange}\times}$ \makebox[0pt][c]{\smash[b]{\rule{0pt}{1.5ex}{\color{white}$+$}}}&
            \cellcolor{LimeGreen}{\color{black}$\checkmark$} \makebox[0pt][c]{\smash[b]{\rule{0pt}{1.5ex}{\color{white}$-$}}}& 
            \cellcolor{LimeGreen}{\color{black}$\checkmark$} \makebox[0pt][c]{\smash[b]{\rule{0pt}{1.5ex}{\color{white}$-$}}} \\ 
            
            \quad 10-19 & \cellcolor{LimeGreen}{\color{black}$\checkmark$} \makebox[0pt][c]{\smash[b]{\rule{0pt}{1.5ex}{\color{white}$-$}}} & \cellcolor{LimeGreen}{\color{black}$\checkmark$} \makebox[0pt][c]{\smash[b]{\rule{0pt}{1.5ex}{\color{white}$-$}}} & \cellcolor{LimeGreen}{\color{black}$\checkmark$} \makebox[0pt][c]{\smash[b]{\rule{0pt}{1.5ex}{\color{white}$-$}}} & \cellcolor{LimeGreen}{\color{black}$\checkmark$} \makebox[0pt][c]{\smash[b]{\rule{0pt}{1.5ex}{\color{white}$-$}}} & \cellcolor{LimeGreen}{\color{black}$\checkmark$} \makebox[0pt][c]{\smash[b]{\rule{0pt}{1.5ex}{\color{white}$-$}}}\\ 
            
            \quad 20-29 & \cellcolor{LimeGreen}{\color{black}$\checkmark_{2/3}$} \makebox[0pt][c]{\smash[b]{\rule{0pt}{1.5ex}{\color{white}$-$}}} & \cellcolor{LimeGreen}{\color{black}$\checkmark$} \makebox[0pt][c]{\smash[b]{\rule{0pt}{1.5ex}{\color{white}$-$}}} & ${\color{RedOrange}\times}$ & \cellcolor{LimeGreen}{\color{black}$\checkmark$} \makebox[0pt][c]{\smash[b]{\rule{0pt}{1.5ex}{\color{white}$-$}}} & \cellcolor{LimeGreen}{\color{black}$\checkmark_{2/3}$} \makebox[0pt][c]{\smash[b]{\rule{0pt}{1.5ex}{\color{white}$-$}}}\\
            
            \quad 30-39 & \cellcolor{LimeGreen}{\color{black}$\checkmark$} \makebox[0pt][c]{\smash[b]{\rule{0pt}{1.5ex}{\color{white}$-$}}} & {\color{black}$\checkmark$} & \cellcolor{LimeGreen}{\color{black}$\checkmark$} \makebox[0pt][c]{\smash[b]{\rule{0pt}{1.5ex}{\color{white}$-$}}} & \cellcolor{LimeGreen}{\color{black}$\checkmark$} \makebox[0pt][c]{\smash[b]{\rule{0pt}{1.5ex}{\color{white}$-$}}} & \cellcolor{LimeGreen}{\color{black}$\checkmark$} \makebox[0pt][c]{\smash[b]{\rule{0pt}{1.5ex}{\color{white}$-$}}} \\ 
            
            \quad 40-49 & {\color{black}$\checkmark$} & \cellcolor{LimeGreen}{\color{black}$\checkmark$} \makebox[0pt][c]{\smash[b]{\rule{0pt}{1.5ex}{\color{white}$-$}}} & \cellcolor{LimeGreen}{\color{black}$\checkmark$} \makebox[0pt][c]{\smash[b]{\rule{0pt}{1.5ex}{\color{white}$-$}}} & \cellcolor{LimeGreen}{\color{black}$\checkmark$} \makebox[0pt][c]{\smash[b]{\rule{0pt}{1.5ex}{\color{white}$-$}}} & \cellcolor{LimeGreen}{\color{black}$\checkmark$} \makebox[0pt][c]{\smash[b]{\rule{0pt}{1.5ex}{\color{white}$-$}}} \\ 
            
            \quad 50-59 & \cellcolor{LimeGreen}{\color{black}$\checkmark$} \makebox[0pt][c]{\smash[b]{\rule{0pt}{1.5ex}{\color{white}$-$}}} & {\color{black}$\checkmark$} & \cellcolor{LimeGreen}{\color{black}$\checkmark$} \makebox[0pt][c]{\smash[b]{\rule{0pt}{1.5ex}{\color{white}$-$}}} & \cellcolor{LimeGreen}{\color{black}$\checkmark$} \makebox[0pt][c]{\smash[b]{\rule{0pt}{1.5ex}{\color{white}$-$}}} & \cellcolor{LimeGreen}{\color{black}$\checkmark$} \makebox[0pt][c]{\smash[b]{\rule{0pt}{1.5ex}{\color{white}$-$}}} \\ 
            
            \quad 60-69 & \cellcolor{LimeGreen}{\color{black}$\checkmark$} \makebox[0pt][c]{\smash[b]{\rule{0pt}{1.5ex}{\color{white}$-$}}} & \cellcolor{LimeGreen}{\color{black}$\checkmark$} \makebox[0pt][c]{\smash[b]{\rule{0pt}{1.5ex}{\color{white}$-$}}} & \cellcolor{LimeGreen}{\color{black}$\checkmark$} \makebox[0pt][c]{\smash[b]{\rule{0pt}{1.5ex}{\color{white}$-$}}} & \cellcolor{LimeGreen}{\color{black}$\checkmark$} \makebox[0pt][c]{\smash[b]{\rule{0pt}{1.5ex}{\color{white}$-$}}} & \cellcolor{LimeGreen}{\color{black}$\checkmark$} \makebox[0pt][c]{\smash[b]{\rule{0pt}{1.5ex}{\color{white}$-$}}} \\ 
            
            \quad 70+ & \cellcolor{LimeGreen}{\color{black}$\checkmark$} \makebox[0pt][c]{\smash[b]{\rule{0pt}{1.5ex}{\color{white}$-$}}} & \cellcolor{LimeGreen}{\color{black}$\checkmark$} \makebox[0pt][c]{\smash[b]{\rule{0pt}{1.5ex}{\color{white}$-$}}} & \cellcolor{LimeGreen}{\color{black}$\checkmark$} \makebox[0pt][c]{\smash[b]{\rule{0pt}{1.5ex}{\color{white}$-$}}} & \cellcolor{LimeGreen}{\color{black}$\checkmark$} \makebox[0pt][c]{\smash[b]{\rule{0pt}{1.5ex}{\color{white}$-$}}} & \cellcolor{LimeGreen}{\color{black}$\checkmark$} \makebox[0pt][c]{\smash[b]{\rule{0pt}{1.5ex}{\color{white}$-$}}} \\ 
            \hline
            \textbf{Fitz. class} & & & & & \\ 
            \quad ~\rom{1} & \cellcolor{LimeGreen}{\color{black}$\checkmark_{2/3}$} \makebox[0pt][c]{\smash[b]{\rule{0pt}{1.5ex}{\color{white}$-$}}} & {\color{black}$\checkmark_{2/3}$} & ${\color{RedOrange}\times}$ & ${\color{RedOrange}\times}$ & ${\color{RedOrange}\times}$ \\
            \quad ~\rom{2} & \cellcolor{pink}${\color{RedOrange}\times}$ & ${\color{RedOrange}\times}$ \makebox[0pt][c]{\smash[b]{\rule{0pt}{1.5ex}{\color{white}$+$}}} & \cellcolor{LimeGreen}{\color{black}$\checkmark_{2/3}$} \makebox[0pt][c]{\smash[b]{\rule{0pt}{1.5ex}{\color{white}$-$}}} & \cellcolor{LimeGreen}{\color{black}$\checkmark_{2/3}$} \makebox[0pt][c]{\smash[b]{\rule{0pt}{1.5ex}{\color{white}$-$}}} & \cellcolor{LimeGreen}{\color{black}$\checkmark$} \makebox[0pt][c]{\smash[b]{\rule{0pt}{1.5ex}{\color{white}$-$}}} \\ 
            \quad ~\rom{3}-~\rom{4} & ${\color{RedOrange}\times}$ & ${\color{RedOrange}\times}$ & ${\color{RedOrange}\times}$ & ${\color{RedOrange}\times}$ & \cellcolor{LimeGreen}{\color{black}$\checkmark_{2/3}$} \makebox[0pt][c]{\smash[b]{\rule{0pt}{1.5ex}{\color{white}$-$}}} \\ 
            \quad ~\rom{5}& ${\color{RedOrange}\times}$ & ${\color{RedOrange}\times}$ & \cellcolor{LimeGreen}{\color{black}$\checkmark$} \makebox[0pt][c]{\smash[b]{\rule{0pt}{1.5ex}{\color{white}$-$}}} & \cellcolor{pink}${\color{RedOrange}\times}$ \makebox[0pt][c]{\smash[b]{\rule{0pt}{1.5ex}{\color{white}$+$}}} & {\color{black}$\checkmark_{2/3}$} \\ 
            \quad ~\rom{6}& ${\color{RedOrange}\times}$ & ${\color{RedOrange}\times}$ & ${\color{RedOrange}\times}$ & ${\color{RedOrange}\times}$ & ${\color{RedOrange}\times}$ \\ 
            \hline
        \end{tabular}
        \label{tab:eoo_gan}
    \end{subtable}
    \hfill
    \begin{subtable}{0.40\textwidth}
        \centering
        \caption{CelebA}
            \label{tab:result_eooCelebA}
        \setlength{\tabcolsep}{6pt} 
        \begin{tabular}{|c|c|c|c|}
            \hline
            \multicolumn{4}{|c|}{\textbf{Sample Size}} \\ 
            \cline{1-4}
            \textbf{100} & \textbf{500} & \textbf{1000} & \textbf{1200} \\ 
            \hline
            & & & \\ 
            \cellcolor{LimeGreen}{\color{black}$\checkmark$} \makebox[0pt][c]{\smash[b]{\rule{0pt}{1.5ex}{\color{white}$-$}}} & \cellcolor{LimeGreen}{\color{black}$\checkmark$} \makebox[0pt][c]{\smash[b]{\rule{0pt}{1.5ex}{\color{white}$-$}}} & \cellcolor{LimeGreen}{\color{black}$\checkmark$} \makebox[0pt][c]{\smash[b]{\rule{0pt}{1.5ex}{\color{white}$-$}}} & \cellcolor{LimeGreen}{\color{black}$\checkmark$} \makebox[0pt][c]{\smash[b]{\rule{0pt}{1.5ex}{\color{white}$-$}}} \\
            \cellcolor{LimeGreen}{\color{black}$\checkmark$} \makebox[0pt][c]{\smash[b]{\rule{0pt}{1.5ex}{\color{white}$-$}}} & \cellcolor{LimeGreen}{\color{black}$\checkmark$} \makebox[0pt][c]{\smash[b]{\rule{0pt}{1.5ex}{\color{white}$-$}}} & \cellcolor{LimeGreen}{\color{black}$\checkmark$} \makebox[0pt][c]{\smash[b]{\rule{0pt}{1.5ex}{\color{white}$-$}}} & \cellcolor{LimeGreen}{\color{black}$\checkmark$} \makebox[0pt][c]{\smash[b]{\rule{0pt}{1.5ex}{\color{white}$-$}}} \\ 
            \cellcolor{LimeGreen}{\color{black}$\checkmark$} \makebox[0pt][c]{\smash[b]{\rule{0pt}{1.5ex}{\color{white}$-$}}} & \cellcolor{LimeGreen}{\color{black}$\checkmark$} \makebox[0pt][c]{\smash[b]{\rule{0pt}{1.5ex}{\color{white}$-$}}} & \cellcolor{LimeGreen}{\color{black}$\checkmark$}  \makebox[0pt][c]{\smash[b]{\rule{0pt}{1.5ex}{\color{white}$-$}}} & \cellcolor{LimeGreen}{\color{black}$\checkmark$} \makebox[0pt][c]{\smash[b]{\rule{0pt}{1.5ex}{\color{white}$-$}}} \\ 
            ${\color{RedOrange}\times}$ & ${\color{RedOrange}\times}$ & ${\color{RedOrange}\times}$ & ${\color{RedOrange}\times}$ \\ 
            \cellcolor{LimeGreen}{\color{black}$\checkmark$} \makebox[0pt][c]{\smash[b]{\rule{0pt}{1.5ex}{\color{white}$-$}}} & {\color{black}$\checkmark_{2/3}$} & \cellcolor{LimeGreen}{\color{black}$\checkmark$} \makebox[0pt][c]{\smash[b]{\rule{0pt}{1.5ex}{\color{white}$-$}}} & \cellcolor{LimeGreen}{\color{black}$\checkmark$} \makebox[0pt][c]{\smash[b]{\rule{0pt}{1.5ex}{\color{white}$-$}}} \\ 
            {\color{black}$\checkmark$} & \cellcolor{LimeGreen}{\color{black}$\checkmark$} \makebox[0pt][c]{\smash[b]{\rule{0pt}{1.5ex}{\color{white}$-$}}} & \cellcolor{LimeGreen}{\color{black}$\checkmark$}  \makebox[0pt][c]{\smash[b]{\rule{0pt}{1.5ex}{\color{white}$-$}}} & \cellcolor{LimeGreen}{\color{black}$\checkmark$} \makebox[0pt][c]{\smash[b]{\rule{0pt}{1.5ex}{\color{white}$-$}}} \\
            \cellcolor{LimeGreen}{\color{black}$\checkmark$} \makebox[0pt][c]{\smash[b]{\rule{0pt}{1.5ex}{\color{white}$-$}}} & \cellcolor{LimeGreen}{\color{black}$\checkmark$} \makebox[0pt][c]{\smash[b]{\rule{0pt}{1.5ex}{\color{white}$-$}}} & \cellcolor{LimeGreen}{\color{black}$\checkmark$}  \makebox[0pt][c]{\smash[b]{\rule{0pt}{1.5ex}{\color{white}$-$}}} & \cellcolor{LimeGreen}{\color{black}$\checkmark$} \makebox[0pt][c]{\smash[b]{\rule{0pt}{1.5ex}{\color{white}$-$}}} \\ 
            \cellcolor{LimeGreen}{\color{black}$\checkmark$} \makebox[0pt][c]{\smash[b]{\rule{0pt}{1.5ex}{\color{white}$-$}}} & \cellcolor{LimeGreen}{\color{black}$\checkmark$} \makebox[0pt][c]{\smash[b]{\rule{0pt}{1.5ex}{\color{white}$-$}}} & \cellcolor{LimeGreen}{\color{black}$\checkmark$}  \makebox[0pt][c]{\smash[b]{\rule{0pt}{1.5ex}{\color{white}$-$}}} & \cellcolor{LimeGreen}{\color{black}$\checkmark$} \makebox[0pt][c]{\smash[b]{\rule{0pt}{1.5ex}{\color{white}$-$}}} \\ 
            \cellcolor{LimeGreen}{\color{black}$\checkmark$} \makebox[0pt][c]{\smash[b]{\rule{0pt}{1.5ex}{\color{white}$-$}}} & \cellcolor{LimeGreen}{\color{black}$\checkmark$} \makebox[0pt][c]{\smash[b]{\rule{0pt}{1.5ex}{\color{white}$-$}}} & \cellcolor{LimeGreen}{\color{black}$\checkmark$}  \makebox[0pt][c]{\smash[b]{\rule{0pt}{1.5ex}{\color{white}$-$}}} & \cellcolor{LimeGreen}{\color{black}$\checkmark$} \makebox[0pt][c]{\smash[b]{\rule{0pt}{1.5ex}{\color{white}$-$}}} \\ 
            \hline
            & & & \\ 
            \cellcolor{pink}{\color{black}$\checkmark_{2/3}$} \makebox[0pt][c]{\smash[b]{\rule{0pt}{1.5ex}{\color{white}$+$}}} & \cellcolor{LimeGreen}{\color{black}$\checkmark_{2/3}$} \makebox[0pt][c]{\smash[b]{\rule{0pt}{1.5ex}{\color{white}$-$}}} & ${\color{RedOrange}\times}$ & ${\color{RedOrange}\times}$ \\ 
            
            ${\color{RedOrange}\times}$ & ${\color{RedOrange}\times}$ & ${\color{RedOrange}\times}$ & ${\color{RedOrange}\times}$ \\ 
            
           \cellcolor{LimeGreen}{\color{black}$\checkmark$} \makebox[0pt][c]{\smash[b]{\rule{0pt}{1.5ex}{\color{white}$-$}}} & \cellcolor{LimeGreen}{\color{black}$\checkmark$} \makebox[0pt][c]{\smash[b]{\rule{0pt}{1.5ex}{\color{white}$-$}}} & \cellcolor{LimeGreen}{\color{black}$\checkmark$} \makebox[0pt][c]{\smash[b]{\rule{0pt}{1.5ex}{\color{white}$-$}}} & {\color{black}$\checkmark$} \\ 
           
            ${\color{RedOrange}\times}$ & ${\color{RedOrange}\times}$ & {\color{black}$\checkmark_{2/3}$} & \cellcolor{LimeGreen}{\color{black}$\checkmark$} \makebox[0pt][c]{\smash[b]{\rule{0pt}{1.5ex}{\color{white}$-$}}} \\ 
            
            \cellcolor{LimeGreen}{\color{black}$\checkmark$} \makebox[0pt][c]{\smash[b]{\rule{0pt}{1.5ex}{\color{white}$-$}}} & ${\color{RedOrange}\times}$ & \cellcolor{LimeGreen}{\color{black}$\checkmark_{2/3}$} \makebox[0pt][c]{\smash[b]{\rule{0pt}{1.5ex}{\color{white}$-$}}} & \cellcolor{LimeGreen}{\color{black}$\checkmark_{2/3}$} \makebox[0pt][c]{\smash[b]{\rule{0pt}{1.5ex}{\color{white}$-$}}} \\ 
            
            \hline
        \end{tabular}
        \label{tab:eoo_celeba}
    \end{subtable}}
\end{table}

\paragraph{Parity test results (1)} 

For the \textit{GAN} dataset, the parity tests of our audit (Table \ref{tab:parity_tests_GAN}) reveal that the observed proportions for gender, age, and Fitzpatrick scale features do not align with the proportions recorded in the general population. For the \textit{CelebA} dataset, the population aged 40 to 59 is the only age group representative of the recorded parameter. 

\paragraph{Equal representation test result (2)}

For the \textit{GAN} dataset, our auditing reveals a strong gender-related bias with ethnicity, for example, a woman is $1.37$ times more likely to be in the Fitzpatrick class \rom{1} compared to a man. Contrariwise, a man is $1.41$ times more likely to be in the Fitzpatrick class \rom{6}  (For all values, see in the Appendix, the Table \ref{tab:Fitzpatrick_selon_genre}).  No gender-related biases with age are present in the \textit{Gan} dataset. 
In \textit{CelebA}, the age group 20-29 is overrepresented among women ($73$\% of women are in this age group against $36$\% for men). There also exist a strong Fitzpatrick-gender bias : while 66\% of women are of Fitzpatrick class \rom{2}, 52\% of men are. On the contrary, men are 12.5 times more likely to be in the Fitzpatrick class \rom{1}.

\section{Conclusion}
\label{sec:conclusion}

We have proposed a new bias auditing method that minimizes the need for manual annotation (requiring between 100 and 1000 annotations) and is robust to errors in automated annotation. When rejecting the null hypothesis, data fluctuations mean that not all statistical tests necessarily yield the same conclusions. For instance, the \(\chi^2\) test appeared more lenient compared to the Wasserstein test. To address this, we consider the majority vote of the results from our three tests. We are also aware that our method tends to accept the null hypothesis ($\mathcal{H}_0$) more readily in the equal representation test. However, we aim to avoid discouraging users from adopting our method due to an excessive number of false rejections. Our primary goal is to encourage users to utilize our tool rather than to achieve a high recall rate (sensitivity).
We emphasize the importance of assessing the representativity of individuals before using an image dataset for training, in order to mitigate potential discrimination.  In the context of the AI Act, which will require companies to certify the compliance of training data for machine learning models, we hope that this audit will serve as a first tool at their disposal.  Our future work will aim to monitor bias in generative models in online mode.

\begin{credits}

\subsubsection{\ackname} 

The authors are partially supported by the AI Interdisciplinary Institute and ANR Regulia , which is funded by the French “Investing for the Future – PIA3” program under the Grant agreement n°ANR-23-IACL-0002. We extend our thanks to Diallo Mohamed, Oumou Hawa Bah and Djiguinée Mamady for their efforts in the labeling process. 

\subsubsection{\discintname}
The authors have no competing interests.
\end{credits}

\newpage


\newpage
\appendix

\section{Fitzpatrick scale definition and demographic distribution}
\label{app:sec:fitz_def}

The Fitzpatrick scale is described as follows, with world proportions recorded in \cite{fc729dd4766245bd817c507b09167924}; \cite{skin_color_lui}:
\begin{itemize}
    \item Phototype I: Very fair skin, freckles, blondes or red hair; prone to always being sunburned without tanning. Approximately 2-5`\% of the world’s population, primarily
in Northern Europe and some populations in Eastern Europe, is affected.
    \item Phototype II: Very fair skin, blondes or light brown hair, light eyes; often sunburned with limited tanning ability. Approximately 10-20\% of the world’s population, primarily in northern and Western Europe, includes some populations in Eastern Europe and North America.
    \item Phototype III: Fair skin, blonde or light brown hair; can gradually tan with variable sensitivity to sunburn. Approximately 20-30\% of the world’s population, is Southern Europe, parts of North America, and mixed populations.
    \item Phototype IV: Olive skin; brown, dark brown, or black hair, dark eyes; rare sunburned; and tans quickly. Approximately 25-35\% of the world’s population, is in
the Mediterranean, Latin America, the Middle East, and parts of South Asia.
    \item Phototype V: Dark skin, dark eyes, dark hair; rarely gets sunburned and tans quickly. Approximately 15-25\% of the world’s population, is in South Asia, the Middle East, North Africa, and some Latino populations.
    \item Phototype VI: Black skin, black hair; rarely prone to sunburn. Approximately 10-15\% of the world’s population, primarily in sub-Saharan Africa and African diaspora populations.
\end{itemize}

\newpage
\section{Difference between the ITA and the Fitzpatrick scale}
\label{app:section:ITA_Fitz}

\begin{figure}
    \centering
    \begin{subfigure}[t]{0.49\textwidth}
        \centering
        \includegraphics[height=1.6in]{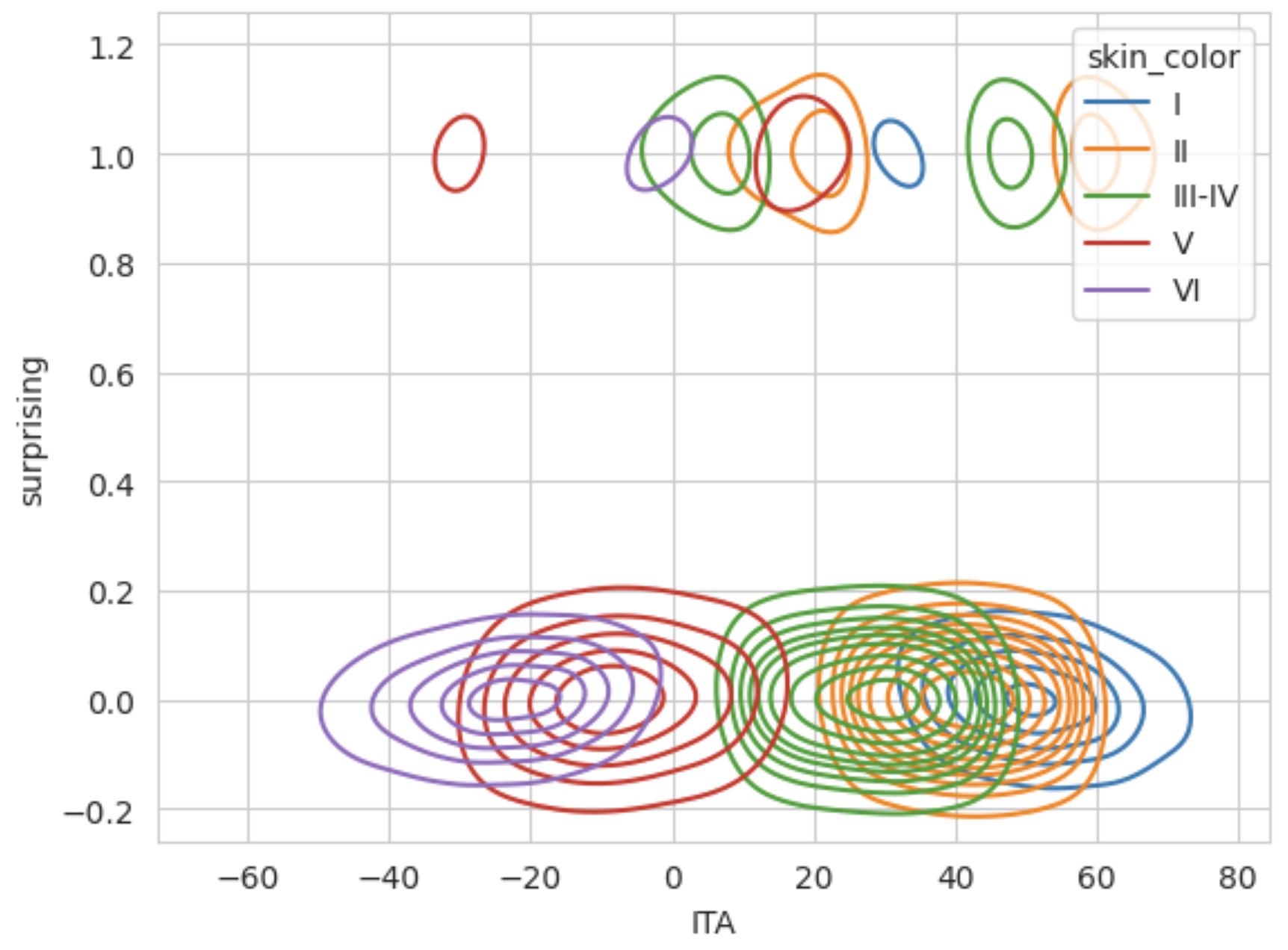}
        \caption{GAN dataset}
        \label{fig:surprisingGAN}
    \end{subfigure}
    \hfill
    \begin{subfigure}[t]{0.49\textwidth}
        \centering
        \includegraphics[height=1.6in]{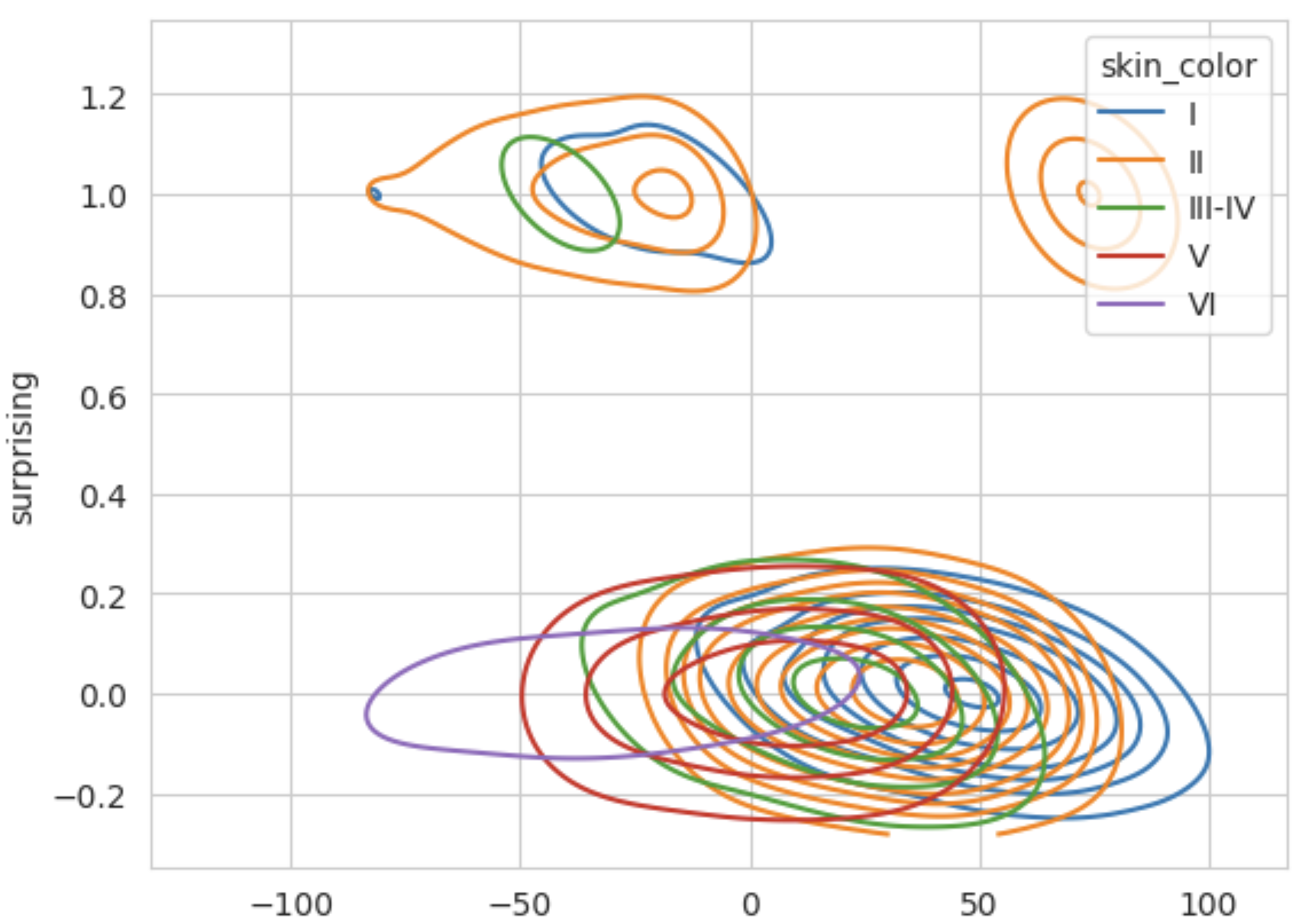}
        
        \caption{Subset of the CelebA dataset}
        \label{fig:surprisingCelebA}
    \end{subfigure}
    \caption{Distribution of ITA given the Fitzpatrick class. A high score of surprising indicates that individuals are located in the tails of the ITA distribution given their Fitzpatrick class.}
    \label{fig:surprising}
\end{figure}

To highlight the differences between the ITA and the Fitzpatrick classes, we estimate mean and standard deviation of ITA distribution for each Fitzpatrick. We then identified extreme observations in the distribution tails by selecting individuals whose ITA values fell below the $2.5\%$ quantile or above the $97.5\%$ quantile of the fitted Gaussian. This approach was inspired from the well-known industrial control chart. We identified individuals whose ITA values were unusually low or high compared to others in the same Fitzpatrick class. For example, we found some individuals with very light skin, manually classified as Fitzpatrick class ~\rom{2} but with an ITA score mostly represented as Fitzpatrick class ~\rom{1}. Due to brown hair, these individuals correspond to Fitzpatrick class ~\rom{2}. This example highlighting discrepancies between ITA-based and Fitzpatrick-based classifications. \\
Moreover, the analysis of our model classifier's accuracy (see Section~\ref{sec:FPforecast}) reveals a slight correlation ($corr= 0.23$) between the network's classification errors and individuals with unusual ITA values compared to the overall group's ITA distribution ($|corr|<0.03$). This correlation suggests that it could even serve as an explainability metric to better understand the reasons behind the classification algorithm's failures.

\begin{figure}
    \centering
    \includegraphics[width=1\linewidth]{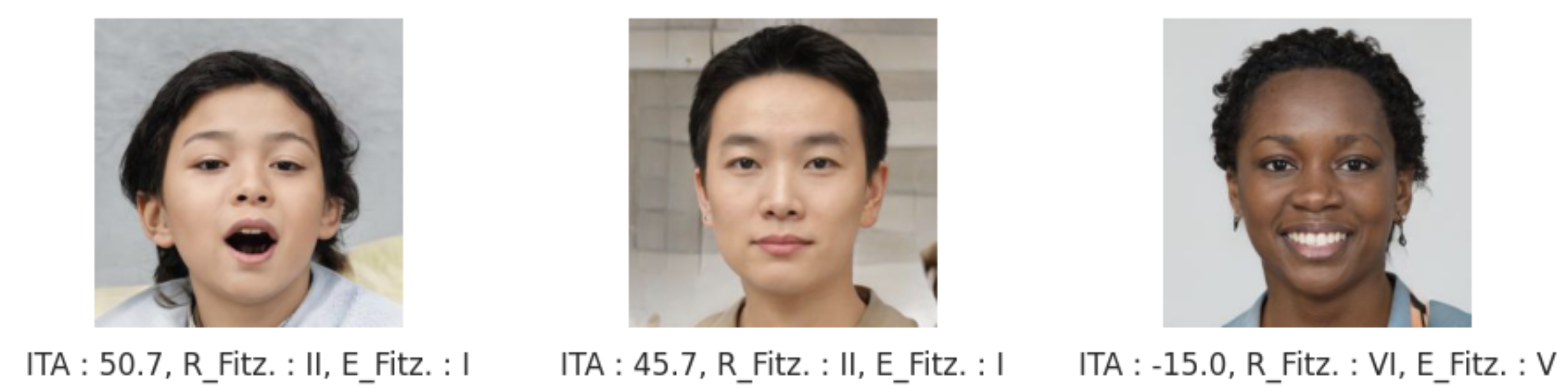}
    \caption{Three examples of individual having a surprising Fitzpatrick class, given their ITA value. Below each picture, we have the ITA values, the Real Fitzpatrick class (according to our annotators), and the Expected Fitzpatrick class according to the ITA.}
    \label{app:fig:Surprising ITA Fitz}
\end{figure}

In Fig. \ref{app:fig:Surprising ITA Fitz}, while skin color of the first two individuals are very light, they are not of the Fitzpatrick I because of their dark eyes and hair. The third individual has a high ITA value for a Fitzpatrick class VI, it might be related to the high luminosity of the picture. 

\section{More information on the methodology}

\subsection{Classification}
\label{app:method:classif}

\begin{figure}[b]
    \centering
    \includegraphics[width=1\linewidth]{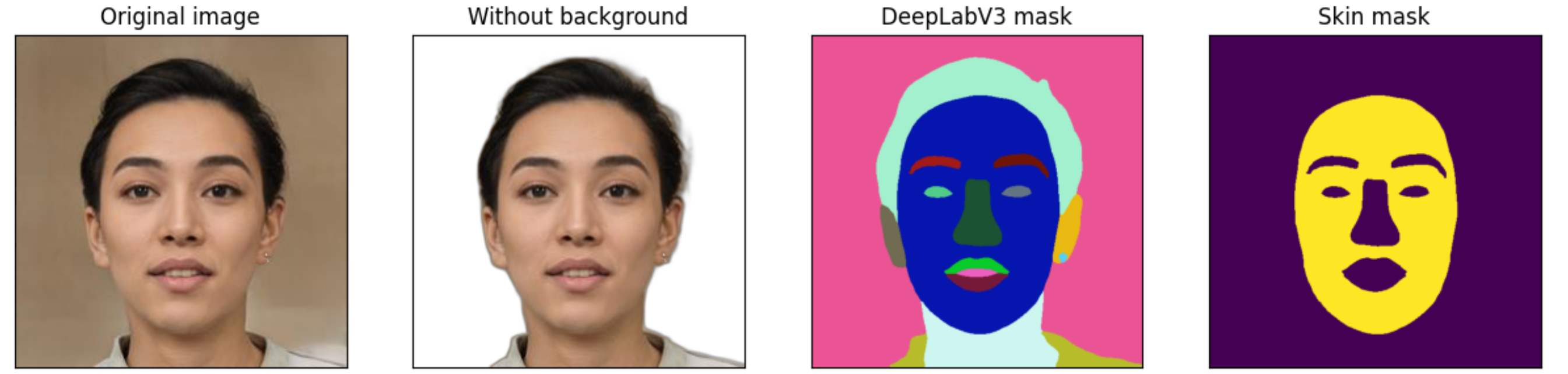}
    \caption{Different mask tried for the Fitzpatrick scale classification}
    \label{fig:mask}
\end{figure}

Other than fine-tuning the Resnet50 and Resnet101, we also experimented with training a ResNet-18 from scratch, as well as using the CNN proposed in \cite{10776291} for classifying Monk Skin Tone. However, both of these approaches produced subpar results.

In addition to the previously mentioned networks, we designed three separate MLP architectures to accommodate different embedding sizes (1,000 or 2,048) and to explore whether incorporating pixel-level information could improve performance. An example of one such “overhead” network is given as follows : 

$$2048 \underset{\text{linear layer + ReLU + BatchNorm}}{\longrightarrow} 256 \underset{\text{concatenate with pixel inf}}{\longrightarrow} 271 \underset{\text{linear layer}}{\longrightarrow} 7$$

For the optimization parameters, we tested 3 optimizers : (1) SGD \cite{SGD}, (2) Adam \cite{Adam} and (3) Schedule free \cite{schedulefree}. Since the schedule-free results outperformed the alternatives, we focus on those in this discussion. After a non-exhaustive hyperparameter search, we selected a learning rate of 0.005.

Because we used pre-trained embeddings from Resnet networks, we could evaluate the performance of our overhead network with fixed embeddings at first, this is a common practice when fine-tuning. This thus allowed us to train for ten epochs our overhead network with 8000 individuals in less than 30 seconds with a T4 GPU. 
We run a grid search across all parameter combinations mentioned previously—such as mask type, ResNet architecture, embedding size, and pixel-level information
We did not use an early stopping method, and kept in memory the validation set and used it to select the best performing epoch, at this epoch for the test result. For each learning configuration, we aggregate the test result of 5 training with a simple mean. This grid-search was done in 2h10 with a T4 GPU.

Having two labeled datasets, and with one bigger than the other, we explored transfer learning as a strategy to reduce the quantity of labeled data needed for accurate Fitzpatrick scale prediction. 
However, the heterogeneity of the two dataset: 
the GAN generated one which was with minimal artifacts (although some images included anomalies like multiple heads) and the CelebA dataset with real-world conditions resulted in mixed outcomes. Models trained on one dataset did not generalize well to the other, and initializing from a model trained on the alternate dataset did not yield any substantial performance gains. Additionally, while accuracy dropped modestly when using a model trained on the real CelebA data and testing on the GAN dataset, the performance decline was notably more severe in the reverse scenario.

\clearpage
\newpage
\subsection{Statistical Test}

\subsubsection{Error-aware Equality of representation Algorithm} \hspace{2.5cm}  \\

\begin{minipage}{\linewidth}%
\begin{algorithm}[H]
\begin{algorithmic}[1]
\State $ acc = \{ \begin{array}{l}
'gender' : \{ 0 : pre\_gender\_men, 1 : pre\_gender\_women \}, \\
'attr' : \{ 0 : acc\_attr\_men, 1 : acc\_attr\_women \}
\end{array} \}$
\State $\text{test\_result} \gets []$
\For{$i \in [0, \text{number\_sim}]$}
    \For{$x_i \in \text{dataset}$}
        \If{$x_i.\text{pred} = \text{True}$}
            \If{$\mathcal{B}(1 - \text{acc}[\text{gender}][x_i.\text{gender}])$}
                \State $x_i.gender \gets 1-x_i.gender$
            \EndIf
            \If{$\mathcal{B}(1 - \text{acc}[\text{attr}][x_i.gender])$}
                \State $\text{sub\_dataset} \gets \{z \in \text{dataset} \mid z.\text{gender} = x_i.\text{gender} \wedge z.\text{pred} = \text{False}\}$
                \State $x_i.\text{attr} \sim \mathcal{U}(\text{sub\_dataset}.\text{attr})$ \Comment{Uniform sampling}
            \EndIf
        \EndIf
    \EndFor
    \State $D_{\text{men}} \gets \{x \in \text{dataset} \mid x.\text{gender} = 0\}$
    \State $D_{\text{women}} \gets \{x \in \text{dataset} \mid x.\text{gender} = 1\}$
    \State $\text{test\_result.append}(\text{test}(D_{\text{men}}, D_{\text{women}}))$
\EndFor
\State $\text{final\_result} \gets \text{median}(\text{test\_result})$
\State \Return $\text{final\_result}$
\end{algorithmic}
\caption{Error-aware Equality of representation Algorithm}   \label{alg:EAER} 
\end{algorithm}
\end{minipage}

\clearpage
\newpage
\subsubsection{Statistical Test choice for continuous-discrete variable: Case study of ITA and Luminescence Across Gender} \hspace{2.5cm}  \\ 

To explore the representation equality for the ITA and the Luminescence two continuous variables across gender, we explore the available statistical test.

\paragraph{Parametric conditions}
The most used test in this setting is the t-test. However, the t-test assumes that the underlying data follows a Gaussian distribution.

\begin{figure}[H]
  \centering
  \includegraphics[width=\linewidth]{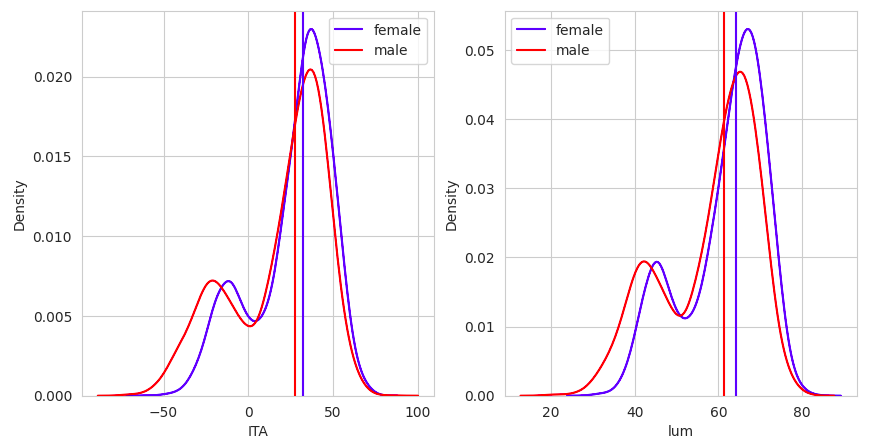}
  \caption{Density plots of ITA and Luminescence values across gender, with vertical lines indicating the medians of each distribution.}
  \label{fig:density_graph_ITA_lum}
\end{figure}

As illustrated in Fig. \ref{fig:density_graph_ITA_lum}, the density plots do not represent Gaussian distributions. To confirm this observation, we conducted the normality test from the \texttt{scipy} library, based on D’Agostino and Pearson’s method. The test yielded p-values below $10^{-100}$, providing strong evidence to reject the null hypothesis ($\mathcal{H}_0$): "the distribution is Gaussian." Consequently, the t-test was deemed unsuitable.

\paragraph{Non-Parametric Testing Approach}

To address the non-Gaussian nature of the data, we considered non-parametric alternatives. The Wilcoxon–Mann–Whitney (WMW) test, a widely recognized non-parametric test, could not be used because of the Behrens-Fisher problem. It highlights that the WMW test, along with other non-parametric localization tests such as the Fisher-Yates-Terry-Hoeding test and the Van der Waerden test, assumes equal dispersion between groups.

Using the Ansari-Bradley test, a non-parametric test for equality of scale, we observed that ITA values across gender violated the dispersion assumption. For Luminescence, the p-value of $0.80$ suggested no such violation. However, both variables failed the Fligner-Killeen test, which examines variance across groups. It is important to note that failing a dispersion test is not conclusive, as these tests themselves assume equal localization parameters —a condition we aimed to evaluate.

Given the uncertainty about dispersion equality, we concluded that the WMW test was inappropriate. Instead, we explored the Fligner-Policello test, a non-parametric alternative that does not assume equal dispersion. However, the Fligner-Policello test requires symmetric distributions around the median. As shown in Fig. \ref{fig:density_graph_ITA_lum}, this symmetry assumption does not hold for our data.

That is why we recommend the Kolmogorov-Smirnov test as well as the Wasserstein test, the two most renowned test which could be applied in our case study (without any assumption on the distribution). 

\section{More results}
\subsection{Classification}

\begin{table}[H]
\centering
\begin{tabular}{|l|l|l|l|l|}
\hline
\textbf{Fitzpatrick class} & \multicolumn{2}{c|}{\textbf{GAN Dataset}} & \multicolumn{2}{c|}{\textbf{CelebA Dataset}} \\  
                            & \textbf{Proportion} & \textbf{Counts}     & \textbf{Proportion} & \textbf{Counts}      \\ \hline
I                   & 8.5\%               & 852                 & 22\%                & 344                  \\ 
II                      & 33.6\%              & 3361                & 56.3\%              & 880                  \\ 
III-IV                    & 32.6\%              & 3360                & 9.7\%               & 152                  \\ 
V                       & 15.2\%              & 1525                & 9.5\%               & 149                  \\ 
VI                    & 10\%                & 1002                & 2.4\%               & 37                   \\ \hline
\end{tabular}
\caption{Fitzpatrick network classification comparison between GAN and CelebA Datasets.}
\label{tab:classif_fitz_comparison}
\end{table}

\begin{table}[H]
    \centering
    \caption{Fitzpatrick class prediction Crosstab (real versus predicted) for GAN and CelebA Datasets}
    \begin{subtable}[t]{0.43\textwidth}
        \centering
        \setlength{\tabcolsep}{4pt}
        \caption{GAN Dataset}
        \begin{tabular}{lccccc}
            \toprule
            \scalebox{0.6}{\diagbox{\textbf{Real}}{\textbf{Predicted}}} & VI & V & III-IV & II & I \\
            \midrule
            VI  & 160 & 46  & 0   & 0   & 0   \\
            V     & 34  & 277 & 17  & 0   & 0   \\
            III-IV  & 0   & 21  & 504 & 103 & 0   \\
            II    & 0   & 0   & 121 & 521 & 21  \\
            I & 0   & 0   & 3   & 45  & 127 \\
            \bottomrule
        \end{tabular}
    \end{subtable}
    \hfill
    \begin{subtable}[t]{0.43\textwidth}
        \centering
        \setlength{\tabcolsep}{4pt}
        \caption{CelebA Dataset}
        \begin{tabular}{lccccc}
            \toprule
            \scalebox{0.6}{\diagbox{\textbf{Real}}{\textbf{Predicted}}} & VI & V & III-IV & II & I \\
            \midrule
            VI  & 5  & 4  & 1  & 1   & 0  \\
            V     & 1  & 9  & 1  & 14  & 0  \\
            III-IV  & 1  & 4  & 1  & 19  & 0  \\
            II    & 3  & 6  & 4  & 141 & 18 \\
            I & 1  & 1  & 0  & 29  & 30 \\
            \bottomrule
        \end{tabular}
    \end{subtable}
\end{table}


\begin{table}[H]
\centering
\caption{FairFace network classification comparison between GAN and CelebA Datasets.}
\begin{tabular}{|l|l|l|l|l|}
\hline
\textbf{FairFace (7 Class)} & \multicolumn{2}{c|}{\textbf{GAN Dataset}} & \multicolumn{2}{c|}{\textbf{CelebA Dataset}}  \\
                            & \textbf{Proportion} & \textbf{Counts}     & \textbf{Proportion} & \textbf{Counts}      \\ \hline
White                      & 21.2\%              & 2102                & 53.9\%              & 839                  \\ 
Middle Eastern             & 6.14\%              & 614                 & 21.1\%              & 328                  \\ 
East Asian                 & 17.34\%             & 1734                & 5.3\%               & 82                   \\ 
Southeast Asian            & 3.43\%              & 343                 & 0.8\%               & 13                   \\ 
Latino Hispanic            & 21.89\%             & 2189                & 6.5\%               & 101                  \\ 
Indian                     & 7.38\%              & 738                 & 2.6\%               & 41                   \\ 
Black                      & 22.8\%              & 2280                & 9.9\%               & 154                  \\ \hline
\end{tabular}
\label{tab:classif_fairface_7_comparison}
\end{table}

\begin{table}[H]
\centering
\caption{Confusion matrix between two age predictions: one supplied by the dataset creator (Gan metadata), the other predictions are from FairFace.}
\setlength{\tabcolsep}{4pt}
\begin{tabular}{|c|c|c|c|c|c|c|c|c|c|}
\hline
\scalebox{0.8}{\diagbox{\textbf{Gan meta}}{\textbf{FairFace }}} & \textbf{0-2} & \textbf{3-9} & \textbf{10-19} & \textbf{20-29} & \textbf{30-39} & \textbf{40-49} & \textbf{50-59} & \textbf{60-69} & \textbf{70+} 
\\ 
\hline
0-2  & 21 & 3   & 0   & 0    & 0    & 0   & 0   & 0   & 0 \\ \hline
3-9  & 5  & 192 & 1   & 0    & 0    & 0   & 0   & 0   & 0 \\ \hline
10-19 & 0  & 238 & 301 & 62   & 0    & 0   & 0   & 0   & 0 \\ \hline
20-29 & 0  & 1   & 329 & 4549 & 230  & 1   & 0   & 0   & 0 \\ \hline
30-39 & 0  & 0   & 2   & 1503 & 1845 & 140 & 0   & 0   & 0 \\ \hline
40-49 & 0  & 0   & 0   & 3    & 105  & 253 & 63  & 0   & 0 \\ \hline
50-59 & 0  & 0   & 0   & 0    & 0    & 5   & 97  & 17  & 0 \\ \hline
60-69 & 0  & 0   & 0   & 0    & 0   & 0 & 9   & 17  & 5 \\ \hline
\end{tabular}
\label{tab:confusion_matrix_predictions_age}
\end{table}

\newpage
\clearpage
\newpage

\subsection{Statistical test}
\subsubsection{Highlighted subgroups by the audit}
\label{app:section:Result_audit}

\begin{table}[b!]
    \centering
    \caption{Distribution of age bracket conditioned by the gender (percentage \%). The Age brackets on which the difference between the two gender was considered significative (with maximum sample size) were put in bold. The real distribution was taken census by \cite{worldbankdata}\cite{ourworlddata}. M : Men, W : Women, T : Total, Dif : $|M - W|$}
    \begin{tabular}{|l|c|c|c|c|c|c|c|c|c|}
    \hline
    \multirow{2}{*}{\textbf{\shortstack{Tranche\\ d'âge}}} 
     & \multicolumn{4}{c|}{\textbf{Gan (10 000)}} & \multicolumn{4}{c|}{\textbf{CelebA (1 500)}} & \textbf{Real}\\ 
            \cline{2-10}  & \textbf{Women} & \textbf{Men} & \textbf{Diff} &  \textbf{Total} & \textbf{Women} & \textbf{Men} & \textbf{Diff} &  \textbf{Total} &  \textbf{Total}\\
            \hline
0-2             & 0  & 0  & 0  & 0.3  & 0 & 0 & 0   & 0   & 2\\
3-9             & 3  & 6  & 2  & 4  & 1 & 0 & 0     & 0.3 & 12\\
10-19           & 5  & 8  & 4  & 6  & 2 & 4 & 2     & 3   & 17\\
\textbf{20-29}  & 69 & 50 & 19 & 61 & \textbf{73} & \textbf{36} & \textbf{37} & 48 & 17\\
30-39           & 19 & 25 & 6  & 22 & 19 & 29 & 9   & 26  & 15\\
40-49           & 2  & 6  & 4  & 4  & 3 & 14 & 11   & 11  & 13\\
50-59           & 1  & 3  & 2  & 2  & 2 & 10 & 8    & 7   & 11\\
60-69           & 0  & 1  & 0  & 0.3  & 0 & 6 & 6   & 4   & 8\\
70+             & 0  & 0  & 0  & 0.05  & 0 & 6 & 6  & 0.1 & 7\\
\hline
    \end{tabular}
    \label{tab:age_selon_genre}
    \vspace{2cm}
\end{table}

\begin{table}[b!]
    \centering
    
    \caption{Fitzpatrick class distribution conditioned on the gender (percentage \%) from manual labelling. The Fitzpatrick classes on which the difference between the two gender was considered significative (with maximum sample size) were put in bold. The real distribution was taken census by \cite{fc729dd4766245bd817c507b09167924} \cite{skin_color_lui}. M : Men, W : Women, T : Total, Dif : $|M - W|$, Max/Min : $\frac{max(M,W)}{min(M,W)}$}
    \begin{tabular}{|l|c|c|c|c|c|c|c|c|c|c|c|}
    \hline
    \multirow{2}{*}{\textbf{\shortstack{Classe de\\ Fitzpatrick}}} 
     & \multicolumn{5}{c|}{\textbf{Gan (10 000)}} & \multicolumn{5}{c|}{\textbf{CelebA (1 500)}} & Real\\ 
            \cline{2-12} & \textbf{W} & \textbf{M} & \textbf{Diff} & \textbf{Max/Min} & \textbf{T} &\textbf{W} & \textbf{M} & \textbf{Diff} & \textbf{Max/Min} & \textbf{T} & \textbf{T}\\
            \hline
    \textbf{\rom{1}} & \textbf{10} & \textbf{7} & 3 & \textbf{1.37} & \textbf{2} 
    & 2 
    & \textbf{31} & \textbf{29} & \textbf{12.5}& 22 & 4 \\
    
    \textbf{\rom{2}} & 32 & 36 & 4 & 1.12  & 34 &
    \textbf{66} & \textbf{52} & \textbf{14} & 1.28 & 56 & 15\\
    
    \rom{3} - \rom{4} & 35 & 30 & 5 & 1.18 & 33 &
    15 & 7 & 7 & 2.03 & 10 & 50\\
    
    \rom{5} & 15 & 16 & 1 & 1.04 & 15 &
    16 & 7 & 9 & 2.34 & 10 & 19\\
    
    \textbf{\rom{6}} & \textbf{9} & \textbf{12} & 3 & \textbf{1.41} & 10 & 
    1 & 3 & 2 & 3.03 & 2 & 12\\

\hline
    \end{tabular}
    \label{tab:Fitzpatrick_selon_genre}
    \vspace{2cm}
\end{table}

\clearpage
\newpage

\subsubsection{Ablation study error aware}

Below are presented the results which would have been obtained by the audit, if we did not use the error aware method (taking into account the inaccuracy of the prediction).

\begin{table}[H]
    \centering
    \caption{Statistical test on the parity \textbf{without taking into account the network accuracy} of Gender ($\mathcal{H}_0$ : $p=0.5$) and the real distribution for the Fitzpatrick Class and Age ($\mathcal{H}_0$ : marginal and global distributions are equivalent). The test used were the Wasserstein test, the Mean test and the \(\chi^2\) test. {$\checkmark$} means none of the tests rejected the hypothesis of being in the corresponding distribution, ${\color{RedOrange}\times}$ means that at least two tests rejected the hypothesis. Colored cell means that the test result are different from the error-aware test, \crule[LimeGreen]{0.3cm}{0.3cm} means that more test rejected $\mathcal{H}_0$.}
    \label{app:tab:parity_tests}
    \scalebox{0.8}{
    \begin{minipage}{0.58\textwidth}
        \centering
        \subcaption{GAN}
          \label{app:tab:parity_tests_GAN}
        \setlength{\tabcolsep}{4pt}
        \begin{tabular}{|l|c|c|c|c|c|}
            \hline
            \multirow{2}{*}{\textbf{\shortstack{Sensitive\\Variable}}} & 
            \multicolumn{5}{c|}{\textbf{Sample Size}} \\ 
            \cline{2-6}
            & \textbf{100} & \textbf{500} & \textbf{1000} & \textbf{3000} & \textbf{8000} \\ 
            \hline
            \textbf{Gender} & ${\color{RedOrange}\times}$ & ${\color{RedOrange}\times}$ & ${\color{RedOrange}\times}$ & ${\color{RedOrange}\times}$ & ${\color{RedOrange}\times}$ \\ 
            \hline
            \textbf{Age} & & & & & \\ 
            \quad 0-2 & \cellcolor{LimeGreen} ${\color{RedOrange}\times}$ & \cellcolor{LimeGreen} ${\color{RedOrange}\times}$ & \cellcolor{LimeGreen} ${\color{RedOrange}\times}$ & \cellcolor{LimeGreen} ${\color{RedOrange}\times}$ & \cellcolor{LimeGreen} ${\color{RedOrange}\times}$ \\ 
            \quad 3-9 & ${\color{RedOrange}\times}$ & ${\color{RedOrange}\times}$ & ${\color{RedOrange}\times}$ & ${\color{RedOrange}\times}$ & ${\color{RedOrange}\times}$ \\ 
            \quad 10-19 & ${\color{RedOrange}\times}$ & ${\color{RedOrange}\times}$ & ${\color{RedOrange}\times}$ & ${\color{RedOrange}\times}$ & ${\color{RedOrange}\times}$ \\ 
            \quad 20-29 & ${\color{RedOrange}\times}$ & ${\color{RedOrange}\times}$ & ${\color{RedOrange}\times}$ & ${\color{RedOrange}\times}$ & ${\color{RedOrange}\times}$ \\ 
            \quad 30-39 & ${\color{RedOrange}\times}$ & ${\color{RedOrange}\times}$ & ${\color{RedOrange}\times}$ & ${\color{RedOrange}\times}$ & ${\color{RedOrange}\times}$ \\ 
            \quad 40-49 & ${\color{RedOrange}\times}$ & ${\color{RedOrange}\times}$ & ${\color{RedOrange}\times}$ & ${\color{RedOrange}\times}$ & ${\color{RedOrange}\times}$ \\ 
            \quad 50-59 & ${\color{RedOrange}\times}$ & ${\color{RedOrange}\times}$ & ${\color{RedOrange}\times}$ & ${\color{RedOrange}\times}$ & ${\color{RedOrange}\times}$ \\ 
            \quad 60-69 & ${\color{RedOrange}\times}$ & ${\color{RedOrange}\times}$ & ${\color{RedOrange}\times}$ & ${\color{RedOrange}\times}$ & ${\color{RedOrange}\times}$ \\ 
            \quad 70+ & ${\color{RedOrange}\times}$ & ${\color{RedOrange}\times}$ & ${\color{RedOrange}\times}$ & ${\color{RedOrange}\times}$ & ${\color{RedOrange}\times}$ \\ 
            \hline
            \textbf{Fitzp. Class} & & & & & \\ 
            \quad ~\rom{1} & ${\color{RedOrange}\times}$ & ${\color{RedOrange}\times}$ & ${\color{RedOrange}\times}$ & ${\color{RedOrange}\times}$ & ${\color{RedOrange}\times}$ \\ 
            \quad ~\rom{2} & ${\color{RedOrange}\times}$ & ${\color{RedOrange}\times}$ & ${\color{RedOrange}\times}$ & ${\color{RedOrange}\times}$ & ${\color{RedOrange}\times}$ \\ 
            \quad ~\rom{3}-~\rom{4} & ${\color{RedOrange}\times}$ & ${\color{RedOrange}\times}$ & ${\color{RedOrange}\times}$ & ${\color{RedOrange}\times}$ & ${\color{RedOrange}\times}$ \\ 
            \quad ~\rom{5}& ${\color{RedOrange}\times}$ & ${\color{RedOrange}\times}$ & ${\color{RedOrange}\times}$ & ${\color{RedOrange}\times}$ & \cellcolor{LimeGreen} {$\checkmark_{2/3}$} \\ 
            \quad ~\rom{6}& ${\color{RedOrange}\times}$ & ${\color{RedOrange}\times}$ & ${\color{RedOrange}\times}$ & ${\color{RedOrange}\times}$ & ${\color{RedOrange}\times}$ \\ 
            \hline
        \end{tabular}
    \end{minipage}
    \hfill
    \begin{minipage}{0.38\textwidth}
        \centering
        \subcaption{CelebA}
          \label{app:tab:parity_tests_CelebA}
        \setlength{\tabcolsep}{6pt}
        \begin{tabular}{|c|c|c|c|}
            \hline
            \multicolumn{4}{|c|}{\textbf{Sample Size}} \\ 
            \cline{1-4}
            \textbf{100} & \textbf{500} & \textbf{1000} & \textbf{1200} \\ 
            \hline
            ${\color{RedOrange}\times}$ & ${\color{RedOrange}\times}$ & ${\color{RedOrange}\times}$ & ${\color{RedOrange}\times}$ \\ 
            \hline
            & & & \\ 
            \cellcolor{LimeGreen} ${\color{RedOrange}\times}$ & \cellcolor{LimeGreen} ${\color{RedOrange}\times}$ & \cellcolor{LimeGreen} ${\color{RedOrange}\times}$ & \cellcolor{LimeGreen} ${\color{RedOrange}\times}$ \\ 
            ${\color{RedOrange}\times}$ & ${\color{RedOrange}\times}$ & ${\color{RedOrange}\times}$ & ${\color{RedOrange}\times}$ \\ 
            ${\color{RedOrange}\times}$ & ${\color{RedOrange}\times}$ & ${\color{RedOrange}\times}$ & ${\color{RedOrange}\times}$ \\ 
            ${\color{RedOrange}\times}$ & ${\color{RedOrange}\times}$ & ${\color{RedOrange}\times}$ & ${\color{RedOrange}\times}$ \\ 
            ${\color{RedOrange}\times}$ & ${\color{RedOrange}\times}$ & ${\color{RedOrange}\times}$ & ${\color{RedOrange}\times}$ \\
            
            \cellcolor{LimeGreen} {$\checkmark_{2/3}$} & \cellcolor{LimeGreen} {$\checkmark_{2/3}$} &
            \cellcolor{LimeGreen} {$\checkmark_{2/3}$} &
            \cellcolor{LimeGreen} {$\checkmark_{2/3}$} \\ 
            
            \cellcolor{LimeGreen} ${\color{RedOrange}\times}$ & \cellcolor{LimeGreen} ${\color{RedOrange}\times}$ & \cellcolor{LimeGreen} ${\color{RedOrange}\times}$ & \cellcolor{LimeGreen} ${\color{RedOrange}\times}$
            \\ 
            
            ${\color{RedOrange}\times}$ & ${\color{RedOrange}\times}$ & ${\color{RedOrange}\times}$ & ${\color{RedOrange}\times}$ \\ 
            
            ${\color{RedOrange}\times}$ & ${\color{RedOrange}\times}$ & ${\color{RedOrange}\times}$ & ${\color{RedOrange}\times}$ \\ 
            \hline
            & & & \\ 
            ${\color{RedOrange}\times}$ & ${\color{RedOrange}\times}$ & ${\color{RedOrange}\times}$ & ${\color{RedOrange}\times}$ \\ 
            ${\color{RedOrange}\times}$ & ${\color{RedOrange}\times}$ & ${\color{RedOrange}\times}$ & ${\color{RedOrange}\times}$ \\ 
            ${\color{RedOrange}\times}$ & ${\color{RedOrange}\times}$ & ${\color{RedOrange}\times}$ & ${\color{RedOrange}\times}$ \\ 
            
            ${\color{RedOrange}\times}$ & ${\color{RedOrange}\times}$ & ${\color{RedOrange}\times}$ & \cellcolor{LimeGreen} {$\checkmark_{2/3}$}\\ 
            
            ${\color{RedOrange}\times}$ & ${\color{RedOrange}\times}$ & ${\color{RedOrange}\times}$ & \cellcolor{LimeGreen} ${\color{RedOrange}\times}$ \\ 
            \hline
        \end{tabular}
    \end{minipage}}
    \vspace{5cm}
\end{table}

\begin{table}[H]
    \centering
    \caption{Equal representation statistical test \textbf{without taking into account the network accuracy} for the Fitzpatrick Class and the Age, with respect to each reflected closest binary gender subgroup. $\mathcal{H}_0$: The Fitzpatrick or Age distribution of the reflected men subgroup is the same as the reflected women subgroup. The tests used were the Wasserstein test, the Mean test, and the \(\chi^2\) test. {$\checkmark$}, {$\checkmark_{2/3}$} and ${\color{RedOrange}\times}$ respectively means that 0, 1 or at least 2 tests rejected $\mathcal{H}_0$. Colored cell means that the test result are different from the error-aware test, \crule[LimeGreen]{0.3cm}{0.3cm} and \crule[pink]{0.3cm}{0.3cm} respectively mean that more test, or respectively less test rejected $\mathcal{H}_0$.
    } 
    \label{app:tab:result_eoo}
    \scalebox{0.8}{\begin{subtable}{0.58\textwidth}
        \centering
        \caption{GAN}
            \label{app:tab:result_eooGAN}
        \setlength{\tabcolsep}{3pt} 
        \begin{tabular}{|l|c|c|c|c|c|}
            \hline
            \multirow{2}{*}{\textbf{\shortstack{Sensitive\\Variable}}} & 
            \multicolumn{5}{c|}{\textbf{Sample Size}} \\ 
            \cline{2-6}
            & \textbf{100} & \textbf{500} & \textbf{1000} & \textbf{3000} & \textbf{8000} \\ 
            \hline
            \textbf{Age} & & & & & \\ 
            \quad 0-2 & \cellcolor{LimeGreen} {$\checkmark_{2/3}$} & \cellcolor{LimeGreen} {$\checkmark_{2/3}$} & \cellcolor{LimeGreen} {$\checkmark_{2/3}$} & \cellcolor{LimeGreen} {$\checkmark_{2/3}$} & \cellcolor{LimeGreen} {$\checkmark_{2/3}$} \\ 
            \quad 3-9 & \cellcolor{LimeGreen} {$\checkmark_{2/3}$} & \cellcolor{LimeGreen} ${\color{RedOrange}\times}$ & \cellcolor{pink} {$\checkmark_{2/3}$} & \cellcolor{LimeGreen} {$\checkmark_{2/3}$} & \cellcolor{LimeGreen} {$\checkmark_{2/3}$} \\ 
            \quad 10-19 & \cellcolor{LimeGreen} {$\checkmark_{2/3}$} & \cellcolor{LimeGreen} {$\checkmark_{2/3}$} & \cellcolor{LimeGreen} {$\checkmark_{2/3}$} & \cellcolor{LimeGreen} {$\checkmark_{2/3}$} & \cellcolor{LimeGreen} {$\checkmark_{2/3}$} \\ 
            \quad 20-29 & \cellcolor{LimeGreen}${\color{RedOrange}\times}$ & \cellcolor{LimeGreen}${\color{RedOrange}\times}$ & ${\color{RedOrange}\times}$ & \cellcolor{LimeGreen} {$\checkmark_{2/3}$} & \cellcolor{LimeGreen} {$\checkmark_{2/3}$} \\ 
            \quad 30-39 & \cellcolor{LimeGreen}${\color{RedOrange}\times}$ & {$\checkmark$} & \cellcolor{LimeGreen}${\color{RedOrange}\times}$ & \cellcolor{LimeGreen}${\color{RedOrange}\times}$ & \cellcolor{LimeGreen} {$\checkmark_{2/3}$} \\ 
            \quad 40-49 & {$\checkmark$} & \cellcolor{LimeGreen} {$\checkmark_{2/3}$} & \cellcolor{LimeGreen} {$\checkmark_{2/3}$} & \cellcolor{LimeGreen} {$\checkmark_{2/3}$} & \cellcolor{LimeGreen} {$\checkmark_{2/3}$} \\ 
            \quad 50-59 & \cellcolor{LimeGreen} {$\checkmark_{2/3}$} & {$\checkmark$} & \cellcolor{LimeGreen} {$\checkmark_{2/3}$} & \cellcolor{LimeGreen} {$\checkmark_{2/3}$} & \cellcolor{LimeGreen} {$\checkmark_{2/3}$} \\ 
            \quad 60-69 & \cellcolor{LimeGreen} {$\checkmark_{2/3}$} & \cellcolor{LimeGreen} {$\checkmark_{2/3}$} & \cellcolor{LimeGreen} {$\checkmark_{2/3}$} & \cellcolor{LimeGreen} {$\checkmark_{2/3}$} & \cellcolor{LimeGreen} {$\checkmark_{2/3}$} \\ 
            \quad 70+ & \cellcolor{LimeGreen} {$\checkmark_{2/3}$} & \cellcolor{LimeGreen} {$\checkmark_{2/3}$} & \cellcolor{LimeGreen} {$\checkmark_{2/3}$} & \cellcolor{LimeGreen} {$\checkmark_{2/3}$} & \cellcolor{LimeGreen} {$\checkmark_{2/3}$} \\ 
            \hline

            \textbf{Fitz. Class} & & & & & \\ 
            \quad ~\rom{1} & 
            \cellcolor{LimeGreen} ${\color{RedOrange}\times}$ &
            {$\checkmark_{2/3}$} & 
            ${\color{RedOrange}\times}$ & 
            ${\color{RedOrange}\times}$ & 
            ${\color{RedOrange}\times}$ \\
            
            \quad ~\rom{2} & 
            \cellcolor{pink} {$\checkmark$} & 
            ${\color{RedOrange}\times}$ & 
            \cellcolor{LimeGreen} ${\color{RedOrange}\times}$ & 
            \cellcolor{LimeGreen} ${\color{RedOrange}\times}$ &
            \cellcolor{LimeGreen} ${\color{RedOrange}\times}$ \\ 
            
            \quad ~\rom{3}-~\rom{4} & 
            ${\color{RedOrange}\times}$ & ${\color{RedOrange}\times}$ & ${\color{RedOrange}\times}$ & ${\color{RedOrange}\times}$ & \cellcolor{LimeGreen} ${\color{RedOrange}\times}$ \\ 
            
            \quad ~\rom{5}& 
            ${\color{RedOrange}\times}$ & ${\color{RedOrange}\times}$ & 
            \cellcolor{LimeGreen}{$\checkmark_{2/3}$} & 
            \cellcolor{pink}{$\checkmark_{2/3}$} & 
            {$\checkmark_{2/3}$} \\ 
            
            \quad ~\rom{6}& 
            ${\color{RedOrange}\times}$ & ${\color{RedOrange}\times}$ & ${\color{RedOrange}\times}$ & ${\color{RedOrange}\times}$ & ${\color{RedOrange}\times}$ \\ 
            \hline
        \end{tabular}
        \label{app:tab:eoo_gan}
    \end{subtable}
    \hfill
    \begin{subtable}{0.38\textwidth}
        \centering
        \caption{CelebA}
            \label{app:tab:result_eooCelebA}
        \setlength{\tabcolsep}{6pt} 
        \begin{tabular}{|c|c|c|c|}
            \hline
            \multicolumn{4}{|c|}{\textbf{Sample Size}} \\ 
            \cline{1-4}
            \textbf{100} & \textbf{500} & \textbf{1000} & \textbf{1200} \\ 
            \hline
            & & & \\ 
\cellcolor{LimeGreen} {$\checkmark_{2/3}$} & \cellcolor{LimeGreen} {$\checkmark_{2/3}$} & \cellcolor{LimeGreen} {$\checkmark_{2/3}$} & \cellcolor{LimeGreen} {$\checkmark_{2/3}$} \\
            
\cellcolor{LimeGreen} {$\checkmark_{2/3}$} & \cellcolor{LimeGreen} {$\checkmark_{2/3}$} & \cellcolor{LimeGreen} {$\checkmark_{2/3}$} & \cellcolor{LimeGreen} {$\checkmark_{2/3}$} \\
            
\cellcolor{LimeGreen} ${\color{RedOrange}\times}$ & \cellcolor{LimeGreen} {$\checkmark_{2/3}$} & \cellcolor{LimeGreen} {$\checkmark_{2/3}$} & \cellcolor{LimeGreen} {$\checkmark_{2/3}$} \\
            
            ${\color{RedOrange}\times}$ & ${\color{RedOrange}\times}$ & ${\color{RedOrange}\times}$ & ${\color{RedOrange}\times}$ \\ 
            
\cellcolor{LimeGreen} ${\color{RedOrange}\times}$ &      
                   {$\checkmark_{2/3}$} & 
\cellcolor{LimeGreen} ${\color{RedOrange}\times}$ & \cellcolor{LimeGreen} {$\checkmark_{2/3}$} \\

{$\checkmark$} & 
\cellcolor{LimeGreen} {$\checkmark_{2/3}$} & \cellcolor{LimeGreen} {$\checkmark_{2/3}$} & \cellcolor{LimeGreen} {$\checkmark_{2/3}$} \\ 
\cellcolor{LimeGreen} {$\checkmark_{2/3}$}& \cellcolor{LimeGreen} {$\checkmark_{2/3}$} & \cellcolor{LimeGreen} {$\checkmark_{2/3}$} & \cellcolor{LimeGreen} {$\checkmark_{2/3}$} \\ 
\cellcolor{LimeGreen} {$\checkmark_{2/3}$}& \cellcolor{LimeGreen} ${\color{RedOrange}\times}$  & \cellcolor{LimeGreen} {$\checkmark_{2/3}$} & \cellcolor{LimeGreen} ${\color{RedOrange}\times}$ \\  

\cellcolor{LimeGreen} {$\checkmark_{2/3}$} & \cellcolor{LimeGreen} {$\checkmark_{2/3}$} & \cellcolor{LimeGreen} {$\checkmark_{2/3}$} & \cellcolor{LimeGreen} {$\checkmark_{2/3}$} \\ 
            \hline
            & & & \\ 
            
            \cellcolor{pink} {$\checkmark$} & \cellcolor{LimeGreen} ${\color{RedOrange}\times}$ & ${\color{RedOrange}\times}$ & ${\color{RedOrange}\times}$ \\ 
            
            ${\color{RedOrange}\times}$ & ${\color{RedOrange}\times}$ & ${\color{RedOrange}\times}$ & ${\color{RedOrange}\times}$ \\ 
            
           \cellcolor{LimeGreen} ${\color{RedOrange}\times}$ & \cellcolor{LimeGreen} {$\checkmark_{2/3}$} & \cellcolor{LimeGreen} {$\checkmark_{2/3}$} & {$\checkmark$} \\ 
           
            ${\color{RedOrange}\times}$ & ${\color{RedOrange}\times}$ & {$\checkmark_{2/3}$} & \cellcolor{LimeGreen} {$\checkmark_{2/3}$}\\ 
            
            \cellcolor{LimeGreen} ${\color{RedOrange}\times}$ & ${\color{RedOrange}\times}$ & \cellcolor{LimeGreen} ${\color{RedOrange}\times}$ & \cellcolor{LimeGreen} ${\color{RedOrange}\times}$ \\ 
            \hline
        \end{tabular}
        \label{app:tab:eoo_celeba}
    \end{subtable}}
\end{table}

Note that even though the general goal of the error-aware method is to prevent false positive tests, the presence of \crule[pink]{0.3cm}{0.3cm} in Table \ref{app:tab:eoo_celeba} is not necessarily a problem. Indeed, as the second correction in our approach brings the predicted distribution closer to the manually labeled distribution, if the predicted distribution is less biased than the manually labeled, then the "correction" can lead to more biases being detected by our method.

\subsection{Disparate Impact}
\label{app:DI}
As stated in the introduction, while the Disparate Impact was not our priority to evaluate the intersectionality between two sensitive variables, it could have been an informative indicator. For instance, using the Disparate Impact alongside with the other three statistical tests results could have been easily argued for. Hence, we provide in this section some result for the equal representation of the Age and the Fitzpatrick class given the reflected gender, for the GAN and CelebA datasets, using the Disparate Impact. We give in Table \ref{app:tab:DI} the upper bound of the $95\%$ confidence interval of the DI as calculated in \cite{besse2018confidenceintervalstestingdisparate} (without using the error-aware method). We observe similarities between the result of the upper bound being over 0.8 and the statistical results presented in the paper.

\begin{table}[H]
    \centering
    \caption{Equal representation analysis via the upper bound of the $95\%$ confidence interval of the Disparate Impact \textbf{without taking into account the network accuracy} for the Fitzpatrick Class and the Age, with respect to each reflected closest binary gender subgroup. 
    } 
    \label{app:tab:DI}
    \scalebox{1}{\begin{subtable}{0.58\textwidth}
        \centering
        \caption{GAN}
            \label{app:tab:DI_GAN}
        \setlength{\tabcolsep}{3pt} 
        \begin{tabular}{|l|c|c|c|c|c|}
            \hline
            \multirow{2}{*}{\textbf{\shortstack{Sensitive\\Variable}}} & 
            \multicolumn{5}{c|}{\textbf{Sample Size}} \\ 
            \cline{2-6}
            & \textbf{100} & \textbf{500} & \textbf{1000} & \textbf{3000} & \textbf{8000} \\ 
            \hline
            \textbf{Age} & & & & & \\ 
            \quad 0-2 & 0.72 & 0.72 & 0.72 & 0.72 & 0.72 \\
            \quad 3-9 & 0.73 & 0.73 & 0.73 & 0.73 & 0.73 \\
            \quad 10-19 & 0.66 & 0.66 & 0.66 & 0.66 & 0.66 \\
            \quad 20-29 & 0.75 & 0.75 & 0.75 & 0.75 & 0.75 \\
            \quad 30-39 & 0.81 & 0.81 & 0.81 & 0.81 & 0.81 \\
            \quad 40-49 & 0.42 & 0.42 & 0.42 & 0.42 & 0.42 \\
            \quad 50-59 & 0.32 & 0.32 & 0.32 & 0.32 & 0.32 \\
            \quad 60-69 & 0.62 & 0.62 & 0.62 & 0.62 & 0.62 \\
            \quad 70+ & 0.00 & 0.00 & 0.00 & 0.00 & 0.00 \\
            \hline

            \textbf{Fitz. Class} & & & & & \\ 
            \quad ~\rom{1} & 0.82 & 1.00 & 0.67 & 0.77 & 0.82 \\
            
            \quad ~\rom{2} & 1.00 & 1.00 & 0.85 & 0.95 & 0.93 \\
            
            \quad ~\rom{3}-~\rom{4} &  0.75 & 0.82 & 0.82 & 0.89 & 0.88 \\
            
            \quad ~\rom{5}& 0.78 & 0.85 & 1.00 & 1.00 & 1.00 \\
            
            \quad ~\rom{6}& 0.27 & 0.81 & 0.67 & 0.71 & 0.77 \\ 
            \hline
        \end{tabular}
    \end{subtable}
    \hfill
    \begin{subtable}{0.38\textwidth}
        \centering
        \caption{CelebA}
            \label{app:tab:DI_CelebA}
        \setlength{\tabcolsep}{6pt} 
        \begin{tabular}{|c|c|c|c|}
            \hline
            \multicolumn{4}{|c|}{\textbf{Sample Size}} \\ 
            \cline{1-4}
            \textbf{100} & \textbf{500} & \textbf{1000} & \textbf{1200} \\ 
            \hline
            & & & \\ 
            1.00 & 1.00 & 1.00 & 1.00 \\
            0.93 & 0.93 & 0.93 & 0.93 \\
            0.84 & 0.84 & 0.84 & 0.84 \\
            0.54 & 0.54 & 0.54 & 0.54 \\
            0.81 & 0.81 & 0.81 & 0.81 \\
            0.32 & 0.32 & 0.32 & 0.32 \\
            0.33 & 0.33 & 0.33 & 0.33 \\
            0.00 & 0.00 & 0.00 & 0.00 \\
            0.00 & 0.00 & 0.00 & 0.00 \\
            
            \hline
            & & & \\ 
            0.97 & 0.15 & 0.19 & 0.13 \\
            1.00 & 0.83 & 0.79 & 0.81 \\
            0.50 & 0.62 & 0.86 & 0.71 \\
            0.94 & 0.62 & 0.53 & 0.58 \\
            0.00 & 0.22 & 0.34 & 0.62 \\
            
            \hline
        \end{tabular}
    \end{subtable}}
\end{table}

\clearpage
\newpage

\section{Additional visualization}

\subsection{Gender \& Fitzpatrick class}
\label{App:vis_gender_fitz}

\begin{figure}
    \centering
    \begin{subfigure}[t]{0.45\textwidth}
        \centering
        \includegraphics[height=1.4in]{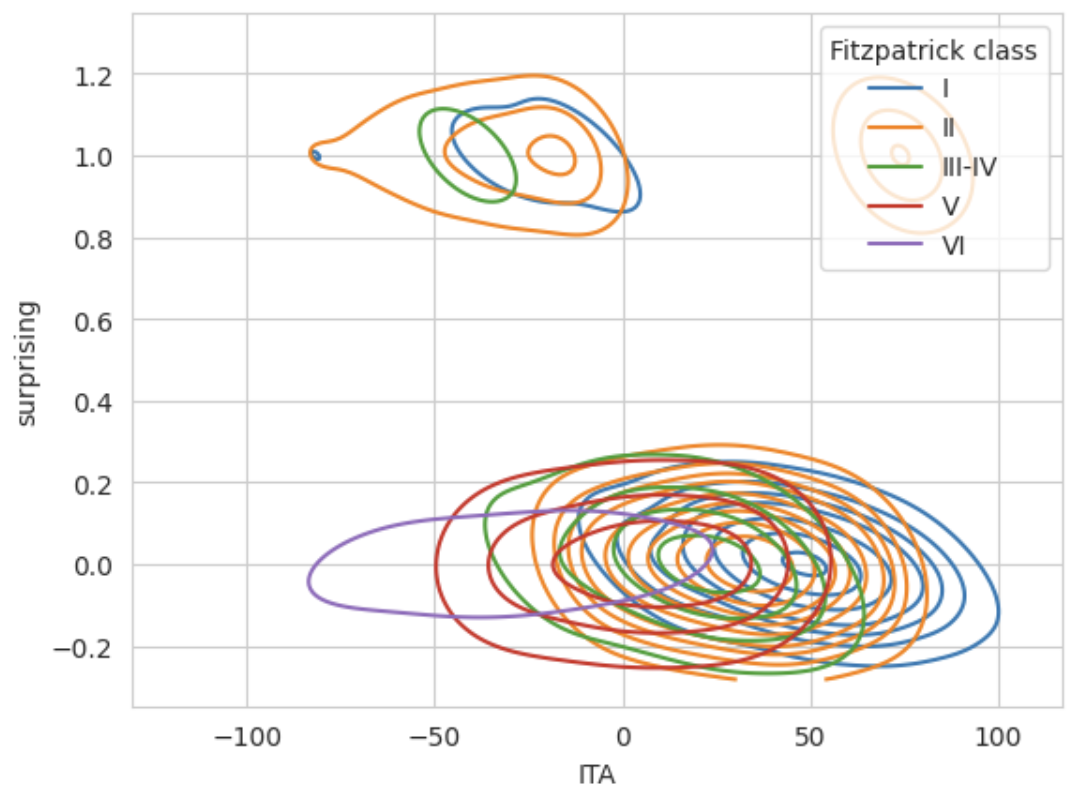}
        \caption{GAN dataset}
        \label{fig:repartitionITAFitzpatrickGAN_gender}
    \end{subfigure}
    \hfill
    \begin{subfigure}[t]{0.45\textwidth}
        \centering
        \includegraphics[height=1.4in]{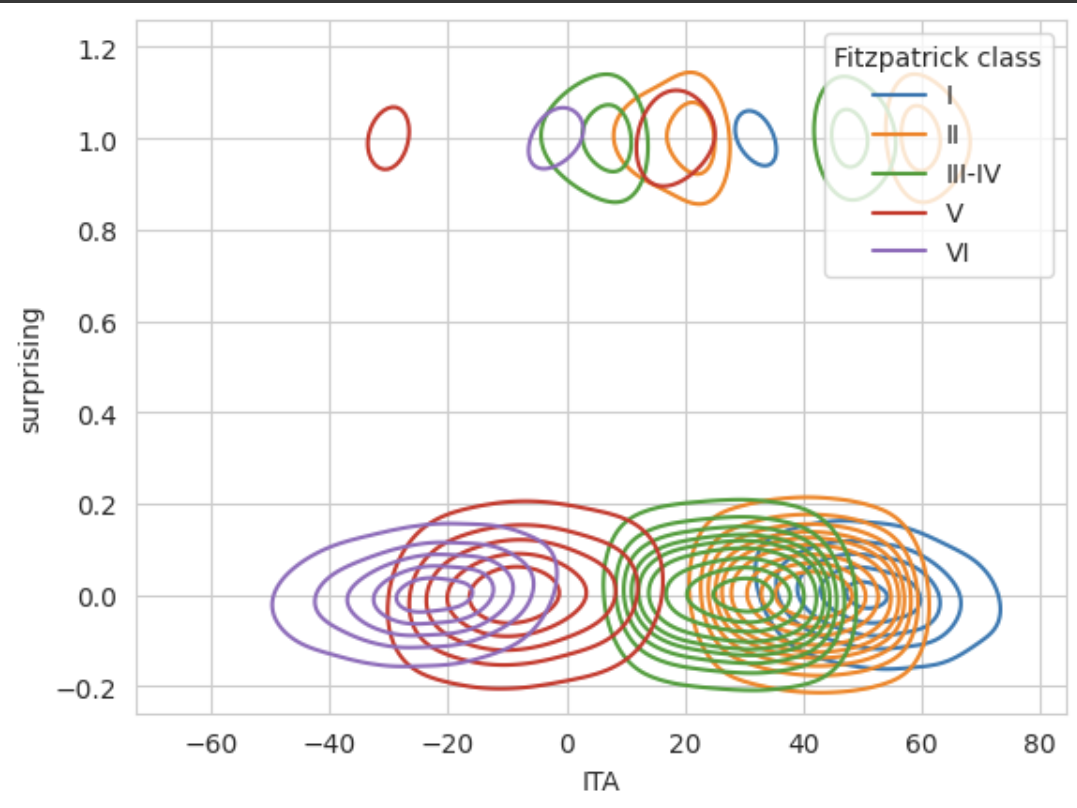}
        \caption{Subset of the CelebA dataset}
        \label{fig:repartitionITAFitzpatrickCeleba_gender}
    \end{subfigure}
    \caption{Probability density of ITA given the Fitzpatrick class and the nearest reflected binary gender, showing a difference between the distribution of reflected binary gender in a subset of the CelebA dataset.}
    \label{fig:repartitionITAFitzpatrick_gender}
\end{figure}

\subsection{FairFace}
\label{app:section:FairFace}

\begin{figure}
    \centering
    \includegraphics[width=0.7\linewidth]{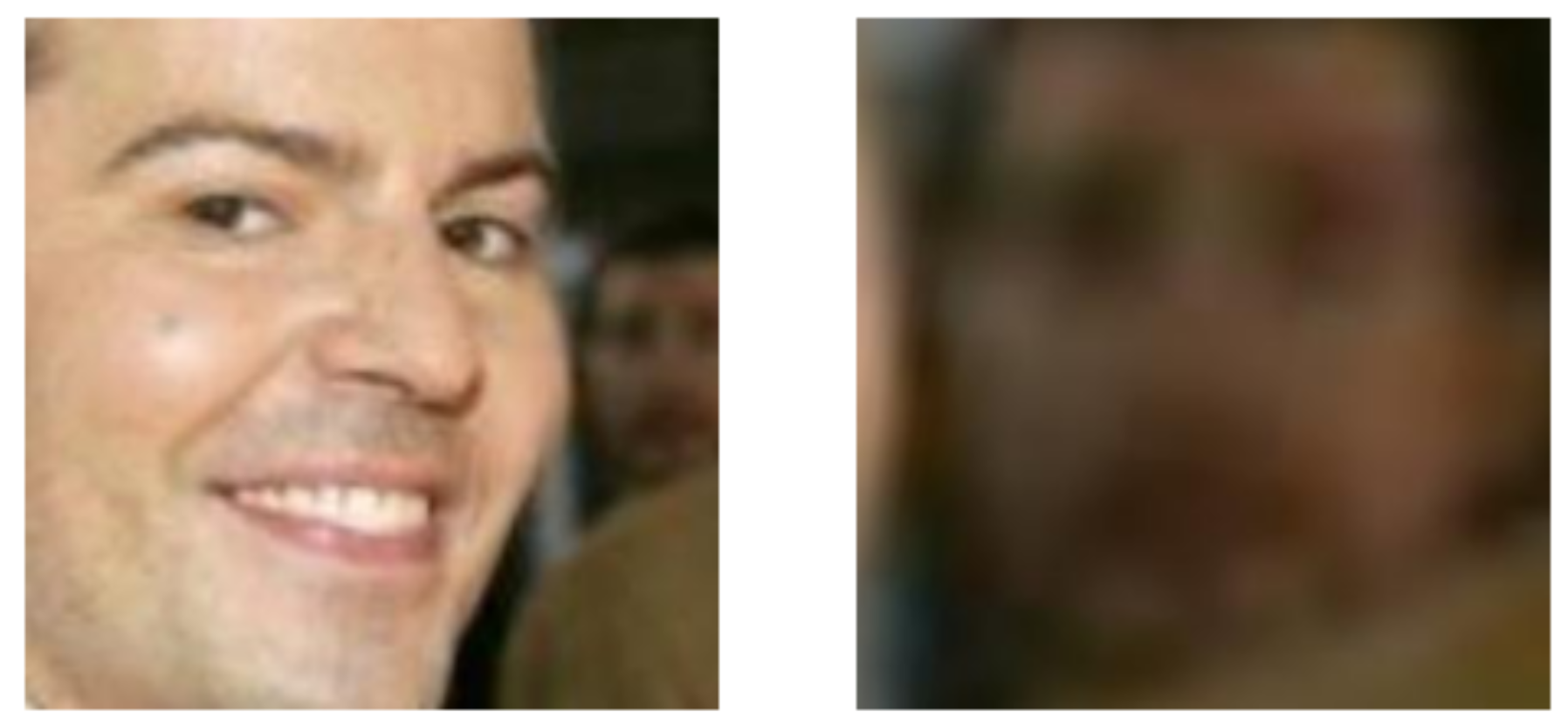}
    \caption{Example of a face in the background detected by FairFace: We observe the right picture being in the background of the left picture. Image from the CelebA dataset.}
    \label{fig:fairfacedetection}
\end{figure}

The FairFace pipeline start by using a face detection algorithm, it led to background face being recognized as seen in Fig.\ref{fig:fairfacedetection}. This can be a little problem if used directly without supervision.
However, FairFace achieved an 97.17\% accuracy on CelebA and an 94.46\% accuracy on GAN for the gender classification, justifying its use even with this undesirable feature (against between 92\% and 93\% for our fine-tuned model for the two datasets).

\end{document}